\documentclass[preprint,5p,times,authoryear]{elsarticle}

\usepackage{mathtools,commath,amssymb,amsthm,bm,mathrsfs}
\usepackage{adjustbox,circledsteps,csquotes,soul}
\usepackage{booktabs,longtable,makecell,tabularray,tabularx,xltabular}
\usepackage{graphicx,float,subcaption}
    \graphicspath{{figures/}}
\usepackage{algorithm,algpseudocode,algorithmicx}
\usepackage{tcolorbox}
    \tcbuselibrary{breakable}
    \tcbset{breakable}
    \newtcolorbox{todo}[1]{title = {#1}, colback=yellow}
\usepackage{comment}
\usepackage{appendix}

\DeclareMathOperator{\dA}{dA}
\DeclareMathOperator{\dx}{\mathrm{d}x}
\DeclareMathOperator{\dy}{\mathrm{d}y}
\DeclareMathOperator{\dz}{\mathrm{d}z}
\DeclareMathOperator{\rect}{rect}

\newcommand{\bX}{\ensuremath{\mathbf{X}}}

\usepackage{nomencl}
    \makenomenclature
    \nomenclature{$A^t$}{Amplitude of the transmitted electric field}
    \nomenclature{$A^r$}{Reflected amplitude}
    \nomenclature{$\mathcal{A}_s$}{Polygonal surface of reflecting plate $s$}
    \nomenclature{$\mathcal{A}^i_p$}{Illuminated region on the final reflecting plate within reflection path $p$}
    \nomenclature{$B$}{Bandwidth}
    \nomenclature{$c$}{Speed of light}
    \nomenclature{$d(\cdot)$}{Normalized vertical position}
    \nomenclature{$d_{\text{max}}$}{Maximum distance allowed between fitted rectangle and reference rectangle}
    \nomenclature{$\epsilon$}{Azimuthal tolerance used for sampling}
    \nomenclature{$E^t$}{Transmitted electric field}
    \nomenclature{$E^r$}{Reflected/received electric field}
    \nomenclature{$f_0$}{Center frequency}
    \nomenclature{$f$}{\Ac{SAR} \ac{ATR} model outputting logits}
    \nomenclature{$\mathcal{G}$}{Subset of samples}
    \nomenclature{$\hat{h}_p$}{Polarization of the incident magnetic field in reflection path $p$ onto the final reflecting plate}
    \nomenclature{$\hat{h}^t$}{Polarization of the transmitted magnetic field}
    \nomenclature{$\vec{H}_p$}{Incident magnetic field in reflection path $p$ onto the final reflecting plate}
    \nomenclature{$I$}{Pixel intensity}
    \nomenclature{$I_{\text{max}}$}{Maximum number of iterations}
    \nomenclature{$\vec{J}$}{Surface density}
    \nomenclature{$\hat{k}_p$}{Direction of propagation of the incident magnetic field in reflection path $p$ onto the final reflecting plate}
    \nomenclature{$\hat{k}^t$}{Direction of propagation of the transmitted magnetic field}
    \nomenclature{$k$}{Phase constant}
    \nomenclature{$K$}{Chirp rate}
    \nomenclature{$l$}{True label of the target}
    \nomenclature{$\mathcal{L}_{CE}(\cdot)$}{Cross-entropy loss function}
    \nomenclature{$\mathcal{L}_{\text{dist}}(\cdot)$}{Distance-based loss}
    \nomenclature{$\mathcal{L}_{\text{pixel}}(\cdot)$}{Pixel-based loss}
    \nomenclature{$m$}{Number of adversarial corner reflectors}
    \nomenclature{$\mathcal{M}^H(\cdot)$}{Measurement in H-polarization}
    \nomenclature{$\mathcal{M}(\cdot)$}{Binary mask function}
    \nomenclature{$N$}{Number of observations}
    \nomenclature{$N^\theta$, $N^\phi$}{Far-field integrals}
    \nomenclature{$\hat{n}_s$}{Unit normal of a reflecting plate $s$}
    \nomenclature{$\vec{p}$}{Position of corner reflector}
    \nomenclature{$\vec{p}^{\,\text{SAR}}$}{Position of the \ac{SAR} platform}
    \nomenclature{$r_0$}{Minimum distance between radar and scene center}
    \nomenclature{$r$}{Distance between the radar and reflector}
    \nomenclature{$R$}{Rectangle defining the bounding box of the target object}
    \nomenclature{$R^{\text{ref}}$}{Reference rectangle}
    \nomenclature{$\text{RDA}(\cdot)$}{Range-Doppler algorithm}
    \nomenclature{$\text{rect}(\cdot)$}{Rectangle function}
    \nomenclature{$t$}{Fast time}
    \nomenclature{$\eta$}{Slow time}
    \nomenclature{$\tau$}{Round-trip time}
    \nomenclature{$T$}{Pulse width}
    \nomenclature{$s$}{Image sample}
    \nomenclature{$\mathcal{S}$}{Clean scene}
    \nomenclature{$\tilde{\mathcal{S}}(\cdot)$}{Perturbation to scene $\mathcal{S}$}
    \nomenclature{$U(\cdot)$}{Uniform function}
    \nomenclature{$v$}{Velocity of the \ac{SAR} platform}
    \nomenclature{$w$, $h$}{Width and height of the observed scene}
    \nomenclature{$x_i$, $y_i$}{Coordinates of corner reflector $i$}
    \nomenclature{$x^R$, $y^R$}{Coordinates of rectangle $R$}
    \nomenclature{$x^\text{ref}$, $y^\text{ref}$}{Coordinates of reference rectangle $R^{\text{ref}}$}
    \nomenclature{$y^R_{\text{min}}, y^R_{\text{max}}$}{Minimum and maximum $y$ coordinates of rectangle $R$}
    \nomenclature{$Z_0$}{Intrinsic impedance of free space}
    \nomenclature{$\alpha$, $\beta$, $\lambda$}{Hyperparameters for fitting the bounding box}
    \nomenclature{$\delta$}{Threshold used to define the reference rectangle}
    \nomenclature{$\Delta$}{Uncertainty in angle of observation}
    \nomenclature{$\theta$}{Incidence angle of corner reflector's boresight}
    \nomenclature{$\theta^a$}{Incidence aspect angle}
    \nomenclature{$\theta'$}{Incidence aspect angle relative to the corner reflector}
    \nomenclature{$\Theta$}{Vector of variables defining the physical perturbation}
    \nomenclature{$\hat{\theta}$}{Estimated incidence angle of observation}
    \nomenclature{$\phi$}{Azimuth angle of corner reflector's boresight}
    \nomenclature{$\phi^a$}{Azimuth aspect angle}
    \nomenclature{$\phi'$}{Azimuth aspect angle relative to the corner reflector}
    \nomenclature{$\Delta \phi^a$}{Azimuth sampling distance}
    \nomenclature{$\hat{\phi}$}{Estimated azimuth angle of observation}

\usepackage[upgrade=true]{acro}
    \acsetup{list/heading=section*, list/template=description}
    \DeclareAcronym{AE}{short=AE, long=adversarial example}
    \DeclareAcronym{AFRL}{short=AFRL, long=Air Force Research Laboratory}
    \DeclareAcronym{ASC}{short=ASC, long=attributed scattering center}
    \DeclareAcronym{ATR}{short=ATR, long=automatic target recognition}
    \DeclareAcronym{BO}{short=BO, long=Bayesian optimization}
    \DeclareAcronym{DARPA}{short=DARPA, long=Defense Advanced Research Projects Agency}
    \DeclareAcronym{DE}{short=DE, long=differential evolution}
    \DeclareAcronym{DNN}{short=DNN, long=deep neural network}
    \DeclareAcronym{DCHUN}{short=DCHUN, long=Deep Convolutional Highway Unit Network}
    \DeclareAcronym{FFT}{short=FFT, long=fast Fourier transform}
    \DeclareAcronym{GAN}{short=GAN, long=generative adversarial network}
    \DeclareAcronym{GEO}{short=GEO, long=geostationary orbit}
    \DeclareAcronym{GO}{short=GO, long=geometrical optics}
    \DeclareAcronym{IFFT}{short=IFFT, long=inverse fast Fourier transform}
    \DeclareAcronym{KLD}{short=KLD, long=Kullback-Leibler divergence}
    \DeclareAcronym{LF2BIM}{short=LF\textsuperscript{2}B-IM, long=Low-Frequency and Feature Bias Iterative Method}
    \DeclareAcronym{LPF}{short=LPF, long=low-pass filter}
    \DeclareAcronym{MIGAA}{short=MIGAA, long=Metasurface Interference-Guided Adversarial Attack}
    \DeclareAcronym{ML}{short=ML, long=machine learning}
    \DeclareAcronym{MSTAR}{short=MSTAR, long=Moving and Stationary Target Acquisition and Recognition}
    \DeclareAcronym{PAA}{short=PAA, long=physical adversarial attack, first-long-format=\itshape}
    \DeclareAcronym{PEC}{short=PEC, long=perfect electric conductor}
    \DeclareAcronym{PO}{short=PO, long=physical optics}
    \DeclareAcronym{PSO}{short=PSO, long=particle swarm optimization}
    \DeclareAcronym{PWFA}{short=PWFA, long=Positively Weighted Feature Attack}
    \DeclareAcronym{RCS}{short=RCS, long=radar cross section}
    \DeclareAcronym{RDA}{short=RDA, long=range-Doppler algorithm}
    \DeclareAcronym{ReLU}{short=ReLU, long=rectified linear unit}
    \DeclareAcronym{RCGA}{short=RCGA, long=real-coded genetic algorithm}
    \DeclareAcronym{SAIPAS}{short=SAIPAS, long=Simulated Aspect-angle-Invariant Physical Adversarial SAR attack, first-long-format=\itshape}
    \DeclareAcronym{SAR}{short=SAR, long=synthetic aperture radar}
    \DeclareAcronym{SAR-PeGA}{short=SAR-PeGA, long=SAR Perturbation Generation Algorithm}
    \DeclareAcronym{SCMA}{short=SCMA, long=Scattering Center Model Attack}
    \DeclareAcronym{SGD}{short=SGD, long=stochastic gradient descent}
    \DeclareAcronym{SMGAA}{short=SMGAA, long=Scattering Model-Guided Adversarial Attack}
    \DeclareAcronym{STARLOS}{short=STARLOS, long=SAR Target Recognition and Location System}
    \DeclareAcronym{SVA}{short=SVA, long=Speckle-Variant Attack}
    \DeclareAcronym{QD}{short=QD, long=quadratic demodulation}

\usepackage{xurl, hyperref}
\usepackage[capitalize,nameinlink]{cleveref} 
    \Crefname{figure}{Fig.}{Figs.}
    \crefname{figure}{fig.}{figs.}
    \crefname{section}{Sec.}{Secs.} 
    \Crefname{section}{Section}{Sections} 
    \Crefname{table}{Table}{Tables} 
    \crefname{table}{Tab.}{Tabs.} 
    
\journal{ISPRS Journal of Photogrammetry and Remote Sensing}

\begin{document}

\begin{frontmatter}

\title{SAIPAS: Simulating aspect-angles-invariant physical adversarial attacks on SAR target recognition models}

\author[eme]{Isar Lemeire\corref{cor1}}
\author[eme]{Yee Wei Law}
\author[eme]{Sang-Heon Lee}
\author[aiml]{William Meakin}
\author[aiml]{Tat-Jun Chin}

\cortext[cor1]{Corresponding author. \textit{E-mail address:} \href{isar.lemeire@adelaide.edu.au}{isar.lemeire@adelaide.edu.au} (I. Lemeire), \href{yeewei.law@adelaide.edu.au}{yeewei.law@adelaide.edu.au} (Y.W. Law), \href{sang-heon.lee@adelaide.edu.au}{sang-heon.lee@adelaide.edu.au} (S. Lee), \href{william.meakin@adelaide.edu.au}{william.meakin@adelaide.edu.au} (W. Meakin), \href{tat-jun.chin@adelaide.edu.au}{tat-jun.chin@adelaide.edu.au} (T. Chin)}

\affiliation[eme]{organization={School of Electrical and Mechanical Engineering, Adelaide University},
                    addressline={Frome Rd},
                    city={Adelaide},
                    postcode={5005},
                    state={South-Australia},
                    country={Australia}}

\affiliation[aiml]{organization={Australian Institute for Machine Learning, Adelaide University},
                    addressline={14 Cnr North Terrace},
                    city={Adelaide},
                    postcode={5005},
                    state={South-Australia},
                    country={Australia}}

\begin{abstract}
\Ac{SAR} enables versatile, all-time, all-weather remote sensing. Coupled with \ac{ATR} leveraging \ac{ML}, \ac{SAR} is empowering a wide range of Earth observation and surveillance applications. However, the surge of attacks based on adversarial perturbations against the \ac{ML} algorithms underpinning \ac{SAR} \ac{ATR} is prompting the need for systematic research into adversarial perturbation mechanisms. Research in this area began in the digital (image) domain and evolved into the (simulated) physical domain, resulting in \acp{PAA} that strategically exploit corner reflectors as attack vectors to evade \ac{ML}-based \ac{ATR}. Existing \acp{PAA} assume that the attacker knows the \ac{SAR} platform’s aspect angles, restricting their applicability to idealized scenarios. We propose the \ac{SAIPAS}, a framework that determines adversarially effective positions and orientations of any given set of reflectors, regardless of their number or size, even when the attacker lacks knowledge of the SAR platform’s aspect angles. This is enabled by rigorous physics-based modeling of the reflected signal and the \ac{SAR} imaging process. To facilitate mapping between image and scene coordinates, we additionally propose a method for generating bounding boxes in densely sampled azimuthal \ac{SAR} images, allowing the target object to serve as a spatial reference. The resulting adversarial configurations offer a clear physical interpretation while maintaining high fooling rates across continuous aspect trajectories under more realistic operational assumptions (69.1\% for AConvNet for a four-reflector white-box attack). This paper has supplementary material available at \href{https://youtu.be/NhpHHQeXaEQ?si=9SRsKGutzPCrxv7J}{https://youtu.be/NhpHHQeXaEQ?si=9SRsKGutzPCrxv7J}, which demonstrates the \ac{SAIPAS}.
\end{abstract}


\end{frontmatter}

\begin{tcolorbox}[
  colback=white,
  colframe=black,
  arc=0mm,
  boxrule=0.5pt,
  title=\bfseries Acronyms,
  coltitle=black,
  colbacktitle=gray!15
]

\let\olddescription\description
\def\description{%
  \olddescription
  \setlength{\itemsep}{-2pt}%
  \setlength{\parsep}{0pt}%
  \setlength{\topsep}{0pt}%
}

\printacronyms[heading=none, template=description]
\end{tcolorbox}

\begin{tcolorbox}[
  colback=white,
  colframe=black,
  arc=0mm,
  boxrule=0.5pt,
  title=\bfseries Nomenclature,
  coltitle=black,
  colbacktitle=gray!15
]

\let\oldsection\section
\renewcommand{\section}[2]{}

\setlength{\nomitemsep}{-3.5pt}

\printnomenclature
\end{tcolorbox}

\section{Introduction}

Synthetic aperture radar (SAR) is a microwave-based active remote-sensing paradigm that improves radar resolution in the azimuth compared to a static radar \cite{jansing2021introduction}. Recent years have witnessed the proliferation of space-based \ac{SAR} systems due to their all-time, all-weather, smoke-penetrating remote sensing capabilities. For example, as of June 2025, Capella Space is operating 7 \ac{SAR} satellites \cite{up402025capella}, while ICEYE is operating 48 \ac{SAR} satellites \cite{iceye2025iceye}. 

A \ac{SAR} system transmits microwave pulses at one location and receives the corresponding echoes at subsequent locations. The transmitted and received signals are then coherently combined (i.e., combined in-phase) to create images of the illuminated terrain \cite{harrison2022introduction}. A wealth of deep learning techniques can readily be leveraged to automatically recognize targets in \ac{SAR} images.
 
From an adversarial perspective, the idea of compromising \ac{SAR} imagery is compelling because \ac{SAR} imagery is generally harder than optical imagery to interpret by human vision. As a result, the potential victim likely relies on algorithms for interpretation, which an attacker may exploit. Moreover, the difficulty of interpreting \ac{SAR} images for humans diminishes the need for visual subtlety. Unlike kinetic and directed energy attacks, the allure of adversarial \ac{ML} attacks (``adversarial attacks'' for short) lies in their stealth and their lack of tendency to escalate into physical conflicts.

More than a decade after Szegedy et al.'s discovery \cite{szegedy2014intriguing}, it is now well known that \acp{DNN} are susceptible to attacks that exploit these networks' lack of robustness in a wide range of data domains, including \ac{SAR}. \emph{Evasion attacks} are a class of adversarial attacks that manipulate test samples, creating so-called \acp{AE}, to evade detection or cause a misclassification by a trained model \cite{nist.ai.100-2e2023}. 
Against \ac{DNN}-based \ac{SAR} \ac{ATR} models, evasion attacks first emerged in the digital domain \cite{du2021adversarial, du2022fast, peng2022scattering, peng2022speckle, chen2023positive, du2023tan, peng2023low, qin2023scma, zhou2023attributed}, where the digital inputs of the targeted \ac{ML} model are directly perturbed. Threats subsequently escalated to the (simulated) physical domain \cite{xia2023sar-pega, yu2023sar, luo2024sar-patt, xie2024migaa, zhang2024physically, ma2025sar-paa, peng2022scattering}, which presents a more concerning threat model \cite{wei2024physical} because the adversary does not require access to the model's digital input pipeline—only the ability to manipulate the physical scene. Most existing works achieve this by strategically deploying corner reflectors as physical attack vectors \cite{xie2024migaa, zhang2024physically, ma2025sar-paa, peng2022scattering}. However, to date, no validated end-to-end physical flight implementations exist in the literature. In this work, we define attacks that constrain perturbations to physically realizable parameters as \acp{PAA} to distinguish them from unconstrained digital pixel perturbations, while noting that all evaluations in the field remain simulation-based.

\begin{figure}[th]
    \centering
    \includegraphics[width=0.55\linewidth]{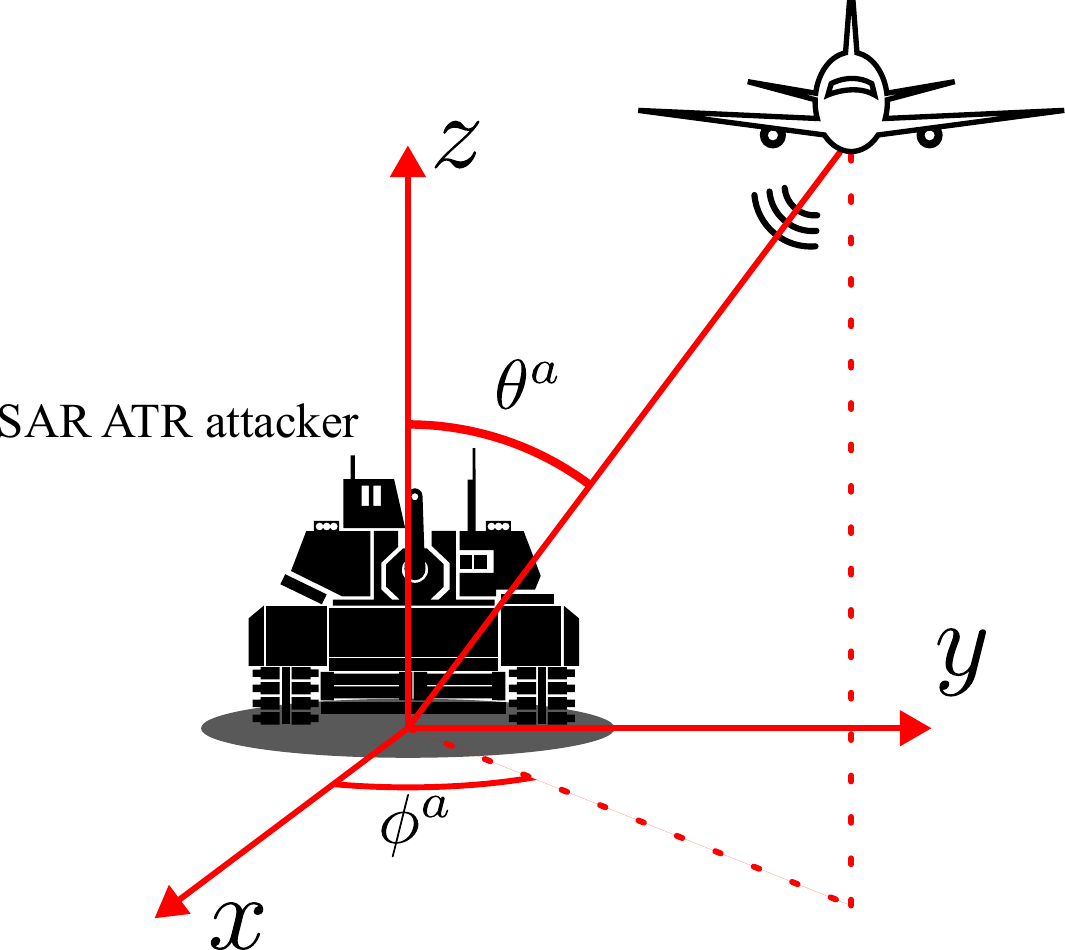}
    \caption{A target object (tank) observed by a \ac{SAR} system from incidence aspect angle $\theta^a$ and azimuth aspect angle $\phi^a$. The attacker attempts to change the physical scene near the target object, to compromise the \ac{SAR} \ac{ATR} model.}
    \label{fig:aspect_angle}
\end{figure}

A knowledge gap in the relevant literature is that prior \acp{PAA} assume the attacker knows the \ac{SAR} system’s aspect angles, shown in \Cref{fig:aspect_angle}, and can perfectly orient the reflectors toward the \ac{SAR} system. These approaches do not model the reflector-induced signal as a function of the aspect angles, which prevents the translation of angular-dependent scattering effects into image-domain perturbations, effectively restricting them to this idealized attacker scenario. Moreover, prior formulations do not explicitly bridge the geometric placement of reflectors in the scene with their induced perturbations in \ac{SAR} image coordinates.

To address these gaps, we propose and analyze a \ac{PAA} against \ac{SAR} \ac{ATR} models, named the \acf{SAIPAS}. The attacker launches an evasion attack by deploying trihedral reflectors on and near the observed target object. Specifically, through this article, we claim the following contributions:

\begin{enumerate}
    \item We propose \ac{SAIPAS}, a simulated \ac{PAA} framework against \ac{DNN}-based \ac{SAR} \ac{ATR} models that improves on prior work by dispensing with the assumption that the \ac{SAR} platform's aspect angles are known to the attacker, motivating its designation as aspect-angles-invariant. Using \ac{SAIPAS}, an attacker possessing a given number of corner reflectors of any dimensions and plate geometry can determine the physical configuration that effectively fools a target model observing the target scene from an unknown angle. The framework is applicable to any target model, as it does not rely on model-specific properties. See Sections~\ref{sec:attack_vectors}-\ref{sec:objective_function} for details.
    
    \item We formulate a physics-driven objective which models the \ac{SAR} observation process, increasing the modeling fidelity compared to prior work. This involves modeling the reflected signal as a function of the aspect angles and the reflectors’ physical properties, and passing the reflected signal through the same measurement and image-focusing operators used in a typical \ac{SAR} processing chain. This formulation enables optimization over a distribution of observations with varying aspect angles. See Sections~\ref{sec:image_formation}-\ref{sec:measure} for details.
    
    \item We propose a method for determining the bounding boxes for a densely sampled azimuthal \ac{SAR} dataset, which leverages the inherent properties of \ac{SAR} images. The bounding boxes enable bridging the geometric placement of reflectors in the scene with their induced perturbations in the image coordinates by using the target object as a consistent spatial reference. See \Cref{sec:bb} for details.
    
    \item We present a comprehensive investigation of optimization strategies, including selection of the optimization algorithm, hyperparameter configuration, and training on limited data. See Sections~\ref{sec:exp_setup}-\ref{sec:finetuning} for details.
\end{enumerate}
As a result, \ac{SAIPAS} demonstrates strong effectiveness across diverse settings (see \Cref{sec:eval_attack}).  Source code will be released on GitHub upon acceptance of this paper.

\section{Related work}

Relevant to our work is the literature on applications of \acp{DNN} to \ac{SAR} \ac{ATR} and adversarial attacks, both digital and physical, targeting \ac{SAR} \ac{ATR}. Adversarial \ac{ML} research in the visible-light domain is relatively well established, and there is no shortage of survey papers \cite{nguyen2024physical, wei2024physical} that adequately summarize the state of the art of adversarial attacks in this domain. As such, the following discussion focuses solely on the \ac{SAR} domain.

\subsection{\acp{DNN} for \ac{SAR} \ac{ATR}}

\ac{SAR} images are typically hard to interpret for humans. When processing a large volume of \ac{SAR} data is time-critical, an \ac{ATR} system becomes necessary. Historically, \ac{ATR} systems relied on traditional model-based or statistical approaches \cite{lang2025recent}. Over the past decade, the rapid advancement of \acp{DNN} has shifted \ac{ATR} research toward data-driven, \ac{ML}-based approaches, which significantly outperform traditional approaches \cite{lang2025recent}.

Despite their promising performance, \acp{DNN} face key challenges in the \ac{SAR} domain, including sensitivity to speckle noise \cite{lang2025recent} and imaging geometry \cite{lang2025recent, zhao2018multi}, as well as a high risk of overfitting due to the limited availability of large high-quality labeled datasets \cite{lang2025recent}. This scarcity arises from the substantial cost of \ac{SAR} data acquisition and the confidentiality associated with many operational datasets \cite{lang2025recent}.

To address these limitations, numerous specialized \acp{DNN} have been proposed for \ac{SAR} \ac{ATR} \cite{zhou2018sar, zhao2018multi, xie2019novel, dong2021global, wang2020sar, shang2018sar, lin2017deep, chen2016target}. Early studies focused on architectural simplification and regularization to mitigate overfitting under limited training data, as seen in models such as AConvNet \cite{chen2016target} and the \ac{DCHUN} \cite{lin2017deep}. Subsequent research shifted toward exploiting richer spatial and contextual representations through hierarchical and multi-branch feature extraction, incorporating multi-scale, multi-stream, and memory-based modules to capture the scattering behavior of targets across varying imaging geometries \cite{zhou2018sar, zhao2018multi, shang2018sar, wang2020sar}. More recent developments extend this trajectory by aiming to capture broader spatial relationships within the scene, by incorporating mechanisms that expand the effective receptive field or aggregate information across distant regions in the image, allowing the network to better represent large-scale structural cues relevant to target shape and orientation \cite{xie2019novel, dong2021global}. In addition to \ac{SAR}-specific models, generic optical classifiers such as AlexNet \\ \cite{krizhevsky2012imagenet}, DenseNet \cite{huang2017densely}, MobileNet \cite{sandler2018mobilenetv2}, and ResNet \cite{he2016deep} have also demonstrated competitive performance on \ac{SAR} \ac{ATR} tasks \cite{li2023comprehensive}. 

This paper adopts AConvNet as the primary architecture for training and evaluating \acp{AE}. In addition, the generated \acp{AE} are tested on standard image classifiers, as outlined in \Cref{sec:atr_models}. These models were chosen as they have been widely employed in prior adversarial machine learning studies against \ac{SAR} \ac{ATR} \cite{peng2022scattering, xie2024migaa, zhang2024physically, ma2025sar-paa}, providing a consistent basis for comparison.

\subsection{Digital attacks on \ac{SAR} \ac{ATR}}

Three major research angles or directions can be observed in the literature: \Circled{1} generation of realistic \acp{AE}, \Circled{2} computational efficiency of the generation process, \Circled{3} transferability of attacks.

\subsubsection{Realistic \acp{AE}} 
Realistic \acp{AE} enforce physical plausibility inside the digital image domain, so the output images look physically coherent, but they do not define a concrete physical deployment mechanism. The rationale for making \acp{AE} realistic is to hamper the detection of artificial perturbations, and is to be differentiated from limiting the $\ell_p$-norm of the perturbations, as constraints based on $\ell_p$-norm or even structural similarity \cite{wang2004image} do not guarantee compliance with the physical laws governing the \ac{SAR} imaging process. Approaches to generating realistic \acp{AE} are either data-driven or model-based.
    
Data-driven approaches rely on the same principle behind generative artificial intelligence \cite{gartner-gen-ai}, i.e., learning from existing artifacts to generate new, realistic artifacts, at scale, that reflect the characteristics of the training data \cite{du2021adversarial}.

Model-based approaches rely on a physics-based model, such as the \ac{ASC} model \cite{potter1997attributed}, for generating artificial \ac{SAR} images. A major benefit of incorporating a physics-based model is that it paves the way for (but not ensure) a physical implementation, i.e., it helps elevate a digital attack to a physical attack. The \ac{ASC} model is widely used to provide guidance on where in a \ac{SAR} image perturbations should be made \cite{qin2023scma, peng2022speckle, zhou2023attributed}.

\subsubsection{Generation efficiency} Some attack schemes focus on the computational efficiency of \ac{AE} generation, for example, by optimizing generative adversarial networks for \ac{AE} generation \cite{du2021adversarial, du2023tan}, or accelerating a traditional digital attack \cite{du2022fast}.

\subsubsection{Transferability} There is growing impetus for making attacks transferable \cite{peng2022speckle, peng2022scattering, peng2023low, chen2023positive, du2023tan}. Among the well-known techniques \cite{gu2024survey}, the \ac{LF2BIM} \cite{peng2023low} capitalizes on the \Circled{1} high-dimensional features of a \ac{SAR} image, accessible from the middle/intermediate layers of a neural network \cite{zhou2018transferable}; \Circled{2} the low-frequency components of a perturbed image to preserve the main structure of the targets in the image; \Circled{3} the gradient calculation algorithm of the translation-invariant attack method \cite{dong2019evading}, together with a Gaussian kernel, to extract low-frequency features. The idea of perturbing high-dimensional features originates in the observation that maximizing the distance between images and their \acp{AE} in the intermediate feature maps enhances attack transferability \cite{zhou2018transferable}.

In the same vein, recent attacks \cite{peng2022speckle, chen2023positive} suppress speckle noise and perturb robust features for transferability. Bernoulli-distributed random masking \cite{he2023improving} can suppress speckle noise \cite{chen2023positive}.
A similar method to perturbing robust features is undoing non-robust perturbations through an ``attenuator'', which is an encoder-decoder network designed to perturb perturbed images to restore correct classifications or predictions \cite{du2023tan}.
Another way of accentuating ``important'' features for transferability can be found in the \ac{PWFA} \cite{chen2023positive}. Maximizing the \ac{KLD} between the positively weighted features of the original image and the positively weighted features of the perturbed image enhances transferability \cite{chen2023positive}. 

The surrogate and target models do not necessarily share similar data distributions. The discrepancy in distributions may diminish transferability between models. To address this domain gap, a surrogate model fine-tuned on synthetic data may be finetuned to be more aligned with target models~\cite{peng2024towards}, or input and feature diversity methods may be deployed \cite{zhang2026towards}.

\subsection{(Simulated) physical attacks on \ac{SAR} \ac{ATR}}

Physical attacks in the optical domain cannot be directly applied to the \ac{SAR} domain due to differences in the physics of the sensing process. \acp{PAA} targeting \ac{SAR} \ac{ATR} systems remain confined to the simulated domain, relying on physical modeling for realism. Although no attacks against \ac{SAR} systems have been physically demonstrated, such methods are broadly categorized as \textit{``physical adversarial attacks''} across the literature. The challenge of evolving digital attacks to the physical domain can be boiled down to the following aspects:
    \subsubsection{Choice of physical attack vectors} For practicality, most physical attack vectors are conceived to be passive, i.e., they reflect \ac{SAR} transmissions and do not produce transmissions. The application of these attack vectors is a form of \emph{passive jamming}, which is the degradation of radar functions by reflecting or absorbing, rather than emitting, electromagnetic waves. The most commonly used passive attack vectors are corner reflectors \cite{peng2022scattering, zhang2024physically, ma2025sar-paa}, as is the case for \ac{SAIPAS}. ``\ac{SAR} stickers'' \cite{yu2023sar, zhang2024physically} and triangular reflective materials \cite{luo2024sar-patt} of uncertain physical properties have also been proposed. Passive but reactive attacks by modulating metasurfaces (e.g., active frequency-selective surfaces, phase-switched screens) in response to received signals have been considered \cite{xia2023sar-pega, xie2024migaa}, but metasurface technologies are still developing.

    \subsubsection{Placement of physical attack vectors} The attack vectors are placed in one of three ways: \Circled{1} only on the target surface \cite{yu2023sar, xia2023sar-pega, luo2024sar-patt}, \Circled{2} only around the target \cite{xie2024migaa, ma2025sar-paa}, \Circled{3} both on and around the target \cite{peng2022scattering, zhang2024physically}, as is the case for \ac{SAIPAS}. In some designs \cite{yu2023sar, zhang2024physically}, placement locations are informed by activation maps, for example, generated with Grad-CAM \cite{selvaraju2017grad-cam}. By placing attack vectors in the shadow regions, the \ac{SMGAA} is not guaranteed to be physically realizable because shadow regions do not reflect \ac{SAR} signals.
    

    \subsubsection{Mapping perturbations in the physical/signal domain to perturbations in the \ac{SAR} digital/image domain} \ac{SMGAA} \cite{peng2022scattering} maps physical scattering to image-domain perturbations using the \ac{ASC} model \cite{akyildiz1999scattering, gerry1999parametric}, but the scattering is generated with a traditional digital attack method instead of a physics-based method. 
    The \ac{SAR-PeGA} \cite{xia2023sar-pega} finds phase modulation sequences for generating echoes, which are then mapped by the \ac{RDA} \cite{raney1994precision} into the image domain.
    In Zhang et al.'s attack \cite{zhang2024physically}, adversarial scatterings are assumed to be strong, allowing a ``simple scattering model'' to be used; \ac{RDA} is then used to map the echo signals into pixels.
    SAR-PATT \cite{luo2024sar-patt} relies on the RaySAR simulator \cite{auer2016raysar}, which however requires material property data that is scarcely available.
    In the \ac{MIGAA} \cite{xie2024migaa}, phase-switched screens are modulated using rectangular waves \cite{wang2018synthetic} and \ac{SAR} images are formed using \ac{RDA}. Nevertheless, physical-to-digital mapping of perturbations is not clearly articulated in every attack design \cite{yu2023sar}. In contrast, the mapping used in \ac{SAIPAS} is entirely physics-based.   
    
    \subsubsection{Optimization formulation} The most common formulation of an attack is maximizing the loss function of the targeted classifier \cite{peng2022scattering, xie2024migaa, ma2025sar-paa}, which is usually the cross-entropy loss; this is the same formulation used in \ac{SAIPAS}. The optimization problem can alternatively be formulated as maximizing the extent of misclassification \cite{yu2023sar, zhang2024physically}, or minimizing a linear combination of negative cross-entropy loss and perturbation \cite{luo2024sar-patt}. In \ac{SAR-PeGA} \cite{xia2023sar-pega}, the optimization problem is finding the phase modulation sequence closest to a Universal Adversarial Perturbation \cite{moosavi-dezfooli2017universal} (a digital attack) pattern.

    \subsubsection{Transferability} Activation maps have been used to guide the placement of physical attack vectors \cite{yu2023sar, zhang2024physically}, following the use of these maps in the RGB domain \cite{wang2021dual}. Besides activation maps, most physical attacks do not apply specific transferability techniques, although they have been evaluated for transferability.

    Digital \acp{AE} in the \ac{SAR} domain have higher fooling rates and are more transferable than those in the optical domain \cite{chen2021empirical}, but no such statement can be made for physical \acp{AE} because of the significant differences in how these physical examples can be implemented in the \ac{SAR} domain and in the optical domain.

\section{Attacker model}

The following attacker model specifies assumptions about the goal, capabilities, and constraints of the attacker, as well as assumptions about the \ac{SAR} system targeted by the attacker, for \ac{SAIPAS}.

\subsection{Assumptions about the attacker}

The attacker seeks to launch an \emph{untargeted} evasion attack, i.e., produce \acp{AE} that cause instances of the target class to be misclassified into any other class \cite{nist.ai.100-2e2023}. Through this untargeted evasion attack, the attacker's ultimate goal is to violate the integrity of the targeted \ac{ML} model \cite{nist.ai.100-2e2023}.

The following assumptions are made about the attacker's knowledge:
\begin{itemize}
    \item The attacker has full knowledge of the \ac{ATR} \ac{ML} model. In \cref{sec:exp}, this assumption is relaxed to evaluate transferability.
    \item The attacker does not know the aspect angles from which the \ac{SAR} system will observe the scene, contrasting with prior attacks that assume the attacker has this knowledge. By removing this requirement, \ac{SAIPAS} adopts a less restrictive attacker model than prior attacks.
    \item The attacker knows the technical specifications of the target \ac{SAR} system.
    
\end{itemize}
Concerning physical implementation, we assume the attacker can place corner reflectors on the ground swath and on the target object and has the time and resources to determine their deployment locations and orientations. 

\subsection{Assumptions about the \ac{SAR} system}\label{sec:assumptions_sar_system}

In this work, the attack formulation assumes a \ac{SAR} system configuration commonly adopted in the literature \cite{li2021adversarial, xia2023sar-pega, ross1998standard}:

\begin{itemize}
    \item The \ac{SAR} system operates in the spotlight mode.
    \item The \ac{SAR} system operates in HH mode, i.e., transmitting and measuring received signals in horizontal polarization.
    \item The \ac{SAR} system uses \ac{QD} to demodulate the received signal.
    \item The \ac{SAR} system uses \ac{RDA} \cite{raney1994precision} as the image formation algorithm.
\end{itemize}

Although the framework is derived for this specific configuration, it is not inherently restricted to it. Alternative acquisition modes, polarization settings, demodulation schemes, or image-focusing algorithms can be accommodated, provided the corresponding signal model and \ac{SAR} imaging pipeline are adapted accordingly.

\section{Proposed attack}\label{sec:proposed_attack}

This section introduces \ac{SAIPAS}, starting with the choice and placement of physical attack vectors, followed by the overall optimization formulation, and the mapping of signal-domain perturbations to the image domain. Finally, the computational complexity of the framework is considered.

\subsection{Choice and placement of physical attack vectors}\label{sec:attack_vectors}

\Ac{SAIPAS} uses trihedral corner reflectors as attack vectors, due to their passive, low-cost nature combined with their ability to produce a bright, localized radar return. The physical perturbation is actuated by $m$ corner reflectors, where each reflector $i$ is parameterized by its position $\vec{p}_i = (x_i, y_i)$, boresight incidence angle $\theta_i$ and azimuth angle $\phi_i$. To ensure maximal azimuthal coverage, their azimuth angles are mutually constrained to be uniformly distributed:
\begin{equation}\label{eq:phi_i}
    \phi_i = \phi_1 + (i-1)\frac{2\pi}{m}.
\end{equation}
Thus, the physical-domain perturbation is parameterized by:
\begin{equation}\label{eq:physical_parameters}
    \Theta = \left[\phi_1, x_1, y_1, \theta_1, \ldots, x_m, y_m, \theta_m\right],
\end{equation}
where $\phi_1 \in \left[0, 2\pi\right] \wedge \forall i: x_i \in \left[-w/2, w/2 \right] \wedge y_i \in \left[-h/2, h/2 \right] \wedge \theta_i \in \left[0, \pi/2\right]  $ where $w$ and $h$ are the scene width and height. Each corner reflector yields a strong return over an azimuthal span of $\pi/2$ \cite{knott2012radar}. While $m \geq 4$ ensures full azimuthal coverage can be achieved, fewer corner reflectors can be used, resulting in reduced efficacy. 

\subsection{Objective function}\label{sec:objective_function}

\begin{figure*}[ht]
    \centering
    \subfloat[]{\includegraphics[width=0.69\linewidth]{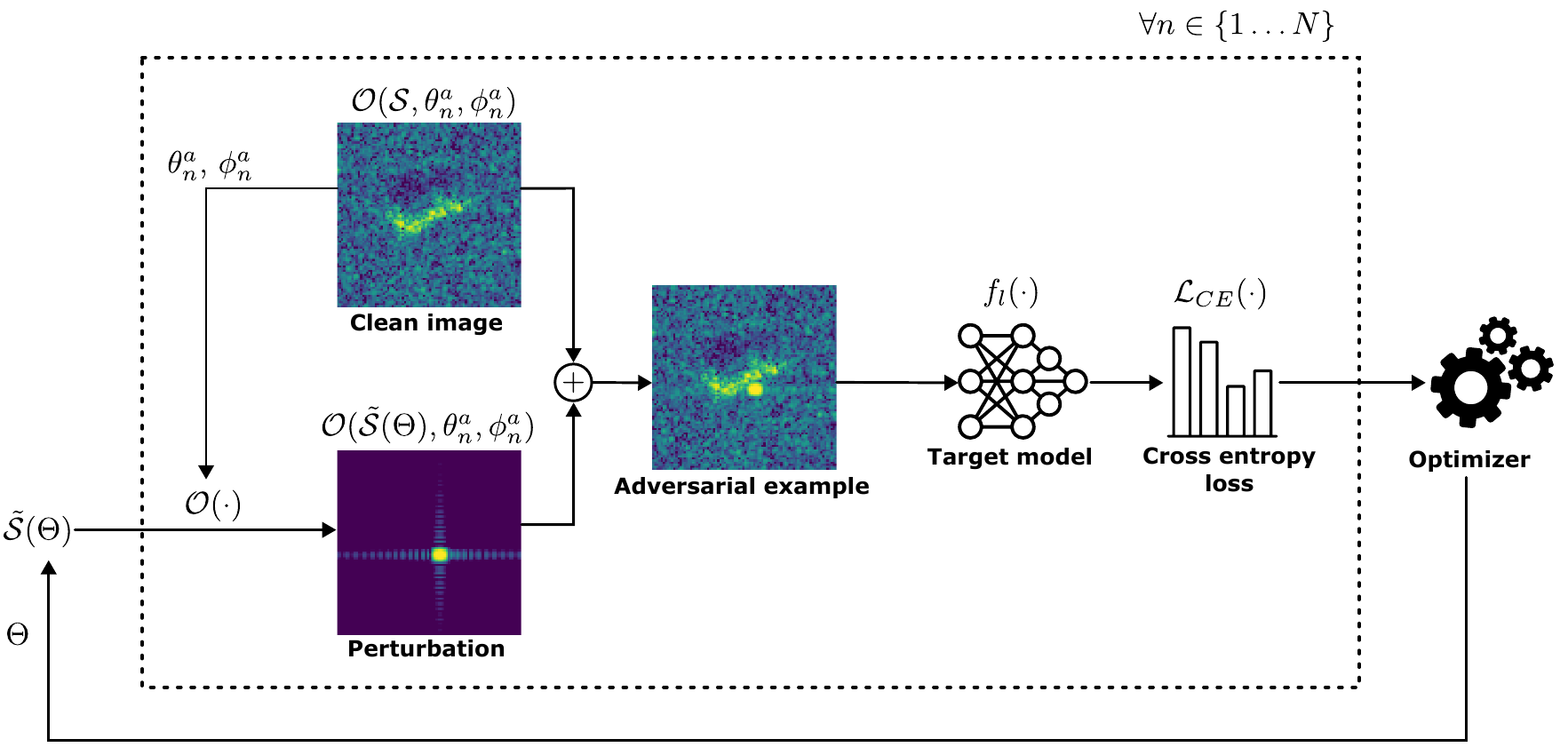}}
    \hfil
    \subfloat[]{\includegraphics[width=0.29\linewidth]{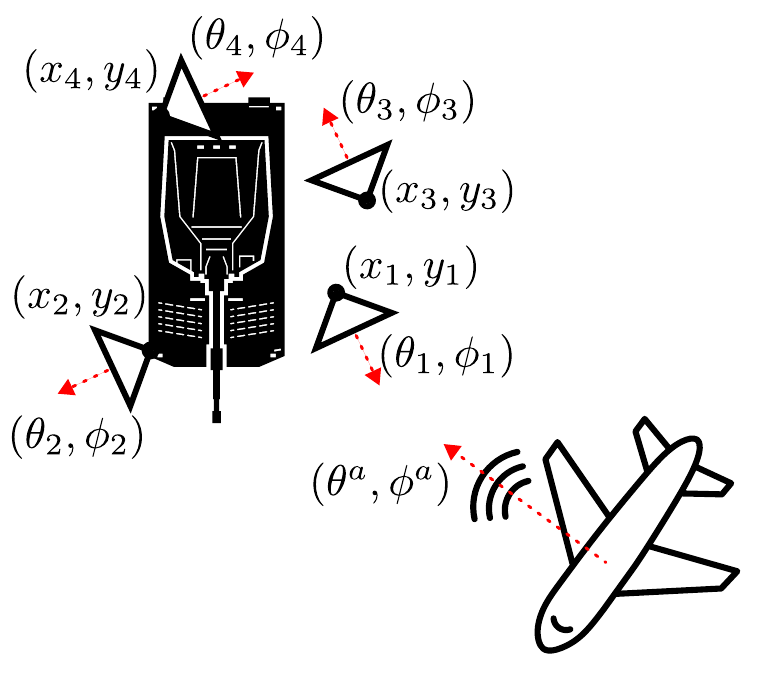}}
    \caption{The strategy of \ac{SAIPAS}: (a) Physical-domain adversarial perturbations actuated by $m$ reflectors are optimized over $N$ observations through Eq.~\eqref{eq:loss}. (b) Top view of a sample reflector configuration, where $m=4$. Each $i$-th reflector is deployed optimally at position $(x_i, y_i)$ with orientation $(\theta_i, \phi_i)$, in a scene observed by a \ac{SAR} platform from angles $(\theta^a, \phi^a)$.}
    \label{fig:attacker_diagram}
\end{figure*}

Let $\mathcal{O}$ denote the \ac{SAR} imaging operator, so $\mathcal{O}\left(\mathcal{S}, \theta^a, \phi^a \right)$ is the image of scene $\mathcal{S}$ observed from incidence aspect angle $\theta^a$ and azimuth aspect angle $\phi^a$, as shown in \Cref{fig:aspect_angle}. \Ac{SAIPAS} seeks to add a physical perturbation $\tilde{\mathcal{S}}(\Theta)$ parameterized by $\Theta$ to the scene $\mathcal{S}$ that causes the target object of class $l$ to be misclassified by \ac{ATR} model $f$ across the entire viewing domain $\theta^a \in \left[0, \pi/2\right] \wedge \phi^a \in \left[0, 2\pi\right]$:
\begin{equation}\label{eq:objective}
    f\left(\mathcal{O}\left(\mathcal{S} + \tilde{\mathcal{S}}(\Theta), \theta^a,  \phi^a\right)\right) \neq l \quad \forall \theta^a \in \left[0, \frac{\pi}{2}\right] \wedge \phi^a \in [0, 2\pi].
\end{equation}
The continuous viewing domain is approximated by a finite set of $N$ \ac{SAR} observations, with $\{(\theta^a_n, \phi^a_n)\}_{n=1,\dots,N}$.

According to the attributed scattering center model, the total high-frequency scattering response can be approximated as a linear superposition of individual scatterers under the \ac{GO} and \ac{PO} regimes~\cite{potter1997attributed, keller1962geometrical}. As detailed in \Cref{sec:physical_model}, a corner reflector can be accurately modelled as a single scatterer center. Furthermore, the image processing is strictly linear~\cite{jansing2021introduction, raney1994precision}. Accordingly, the perturbed image is approximated by the superposition of the clean image and the perturbation:
\begin{equation}\label{eq:adversarial_example}
    \mathcal{O}\left(\mathcal{S} + \tilde{\mathcal{S}}(\Theta), \theta^a,  \phi^a\right) \approx  \mathcal{O}\left(\mathcal{S}, \theta^a,  \phi^a\right) + \mathcal{O}\left(\tilde{\mathcal{S}}(\Theta), \theta^a,  \phi^a\right).
\end{equation} 

Because directly satisfying the misclassification condition in \cref{eq:objective} is non-continuous, we follow standard adversarial \ac{ML} practices and relax the objective using the continuous cross-entropy loss $\mathcal{L}_\text{CE}$ as a surrogate. The adversarial parameters $\Theta$ are determined by solving:
\begin{equation}\label{eq:loss}
\begin{aligned}
      \max_{\substack{\Theta}}\; & \frac{1}{N} \sum_{n = 1} ^ N \mathcal{L}_{CE}\left( f \left( \mathcal{O}\left(\mathcal{S}, \theta^a_n,  \phi^a_n\right) + \mathcal{O}\left(\tilde{\mathcal{S}}(\Theta), \theta^a_n,  \phi^a_n \right) \right), l\right), \\
    \text{s.t.}\; & \forall i \in \{1, \ldots, m \}: \phi_1 \in \left[0, 2\pi\right] \wedge  \theta_i \in \left[0, \frac{\pi}{2}\right] \\ & \wedge x_i \in \left[-\frac{w}{2}, \frac{w}{2} \right] \wedge y_i \in \left[-\frac{h}{2}, \frac{h}{2} \right].
\end{aligned}
\end{equation}
The overall attack architecture is illustrated in \Cref{fig:attacker_diagram}.

\subsection{\Ac{SAR} Imaging process}\label{sec:image_formation}

Given the assumptions specified in \Cref{sec:assumptions_sar_system}, creating a \ac{SAR} image involves the following sequential steps:
\begin{enumerate}
    \item As the \ac{SAR} platform traverses a predefined flight path, it transmits a sequence of identical signal pulses $E^t(t)$, expressed as a function of fast time $t$, toward the scene.
    \item The transmitted signals are reflected by objects within the scene, where each pulse yields a different reflection at a different point in slow time $\eta$, resulting in a backscattered signal $E^r(t, \eta)$.
    \item The reflected signal is measured, demodulated using \ac{QD}, and sampled over fast and slow time, resulting in a two-dimensional matrix.
    \item The demodulated signal is focused into a \ac{SAR} image using \ac{RDA}.
\end{enumerate}
Thus, the image produced by the physically perturbed scene is computed by modeling the \ac{SAR} imaging process:
\begin{equation}
\mathcal{O}\left(\tilde{\mathcal{S}}(\Theta), \theta^a_n,  \phi^a_n\right) = \text{RDA} \left( \text{QD} \left( \mathcal{M}\left( \sum_{i = 1}^m E^r_{i, n}(t, \eta) \right) \right)  \right),
\end{equation}
where $E^r_{i, n}(t, \eta)$ is the reflected signal from the $i$-th corner reflector for the $n$-th observation (with corresponding aspect angles $\theta^a_n,  \phi^a_n$), $\mathcal{M}(\cdot)$ denotes the measuring operator, $\text{QD}(\cdot)$ denotes the quadratic demodulation operator and $\text{RDA}(\cdot)$ denotes the range-Doppler algorithm operator.

\subsection{Physics-based reflection model}\label{sec:physical_model}

The following shows a derivation of an expression for $E^r_{i, n}(t, \eta)$ for a single reflector $i$ and observation $n$. For brevity, $(i,n)$  are omitted. Accordingly, $(\theta, \phi)$ and $\vec{p}$ denote the orientation and position of said corner reflector, while $\theta^a, \phi^a$ denote the aspect angles.

Each transmitted pulse $E^t(t)$ is a linear frequency-modulated waveform, commonly called a \emph{chirp} \cite{jansing2021introduction}:
\begin{equation}\label{eq:chirp}
    E^t(t) = A^t \rect \left(\frac{t}{T}\right) \cos\del{2\pi f_0 t + \pi K t^2},
\end{equation}
where $A^t$ is the amplitude of the transmitted pulse, $T$ is the pulse duration, $f_0$ is the center frequency, $K$ is the chirp rate, and $\rect$ is the rectangle function.

\begin{figure}[ht]
    \centering
    \includegraphics[width=0.4\linewidth]{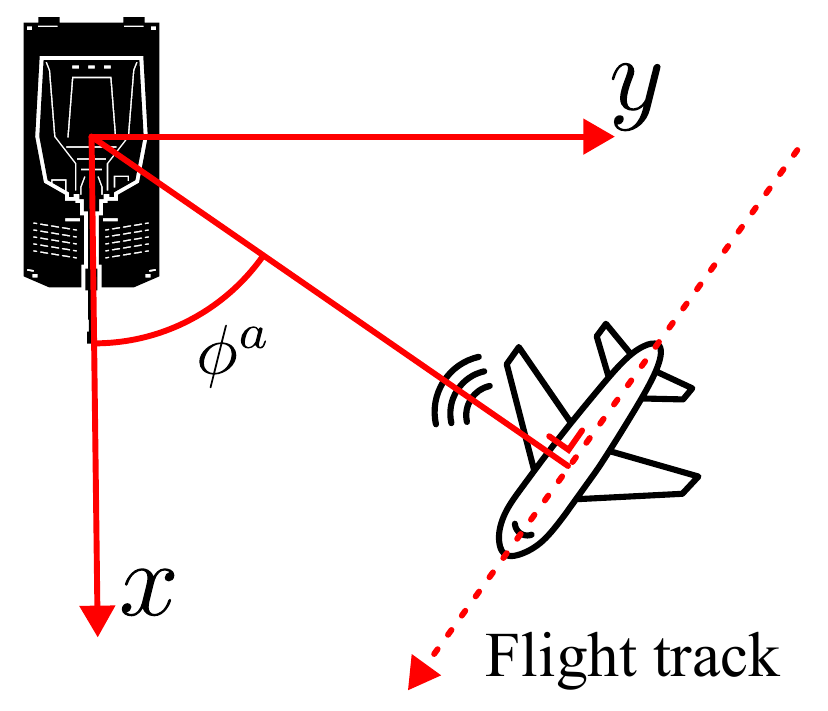}
    \caption{Top view of a target object (e.g., tank) observed by a \ac{SAR} system from azimuth aspect angle $\phi^a$. The flight track is assumed to be perpendicular to the line of sight.}
    \label{fig:azimuth_angle}
    \centering\includegraphics[width=0.4\linewidth]{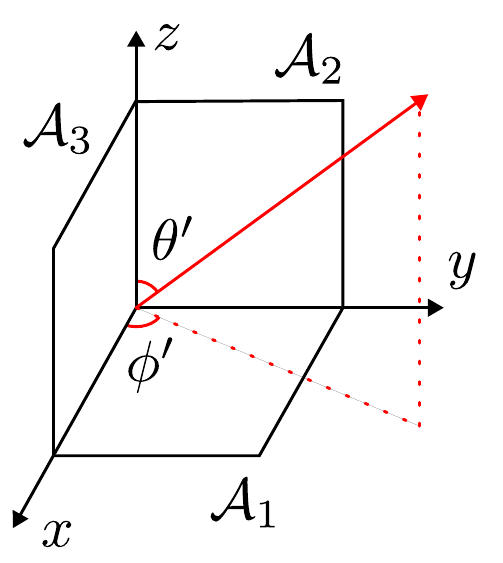}
    \caption{Numbered plates $\mathcal{A}_1$, $\mathcal{A}_2$, $\mathcal{A}_3$ of a square trihedral corner reflector. The coordinate frame is aligned with the plates, and aspect angles $\theta'$, $\phi'$ are defined in this frame.}
    \label{fig:surfaceNames}
\end{figure}

To model the backscatter of the corner reflector, we build on top of a method developed by \cite{polycarpou1995radar}, which has been shown to produce predictions closely aligned with real measurements. The reflected electric field can be expressed in terms of the far-field spherical components $E^\theta(\eta)$, $E^\phi(\eta)$ \cite{balanis2012advanced}:
\begin{align}\label{Eq:amplitude}
    E^r(t, \eta) &\approx E^\theta(t, \eta) \hat{\theta} + E^\phi(t, \eta)\hat{\phi},
\end{align}
where $\hat{\theta}$ and $\hat{\phi}$ are unit vectors along the incidence ($\theta$) and azimuth ($\phi$) spherical coordinate directions, respectively. Each spherical component is a summation of fifteen reflection components, each caused by a different reflection path $p$ \cite{polycarpou1995radar}:
\begin{equation}\label{Eq:reflections}
\begin{aligned}
E^{\theta}(t,\eta) \approx& \sum_{p \in \mathcal{P}} E^{\theta}_{p}(t, \eta),  \quad
E^{\phi}(t, \eta) \approx \sum_{p \in \mathcal{P}} E^{\phi}_{p}(t, \eta), \\
\mathcal{P} = & \;\{
1, 2, 3, \\
 &12, 21, 13, 31, 23, 32, \\
 &123, 132, 213, 231, 312, 321
\},
\end{aligned}
\end{equation}
where, assuming $\mathcal{A}_1$, $\mathcal{A}_2$, $\mathcal{A}_3$ denote the reflector plates of the reflector as illustrated on \Cref{fig:surfaceNames}, $p = 1$  corresponds to a single bounce off surface $\mathcal{A}_1$, $p = 12$ corresponds to the reflection path with a double bounce, first off surface $\mathcal{A}_1$ then surface $\mathcal{A}_2$, and $p = 123$ corresponds to the reflection path with a triple bounce in the order of surfaces $\mathcal{A}_1$, $\mathcal{A}_2$, and $\mathcal{A}_3$. The remaining terms follow the same naming convention. Different reflector geometries can be incorporated into the model by varying the polygonal description of $\mathcal{A}_1$, $\mathcal{A}_2$, $\mathcal{A}_3$.

To simplify the analysis, the aspect angles are expressed relative to the corner reflector's boresight \cite{knott2012radar}:
\begin{equation}
\phi' = \phi^a - \left(\phi - \frac{\pi}{4}\right), \quad \theta' = \theta^a - \left(\theta - \arctan(\sqrt{2})\right),
\end{equation}
aligning the coordinate frame so that each plate of the trihedral aligns with a coordinate plane, as illustrated in \Cref{fig:surfaceNames}. Only $\phi', \theta' \in [0, \pi/2]$ are considered, as scattering outside this range is negligible \cite{knott2012radar}. Additionally, since aspect angle variation across the observation is negligible, $\theta'$ and $\phi'$ are held constant across slow time at their trajectory-center values.

The backscattered far-field components caused by the reflection path $p$ are:\cite{balanis2012advanced, polycarpou1995radar}:
\begin{equation}\label{Eq:spherical_components}
\begin{aligned}
    E^{\theta}_{p}(t, \eta) &\approx \frac{-j Z_0 N^{\theta}_{p}}{2\lambda} \frac{e^{j  \psi(t, \eta) }}{r(\eta)}, \\
    E^{\phi}_{p}(t, \eta) &\approx \frac{-j Z_0 N^{\phi}_{p}}{2 \lambda} \frac{e^{j \psi(t, \eta)}}{r(\eta)}, \\
\end{aligned}
\end{equation}
where $N^{\theta}_p$ and $N^{\phi}_p$ are the far-field integrals, and $\psi(t, \eta)$ the phase, which is identical to the phase of the transmitted signal defined in \cref{eq:chirp}, but delayed by round trip time $\tau$:
\begin{equation}
    \psi(t, \eta) = 2\pi f_0(t - \tau(\eta)) + \pi K(t - \tau(\eta))^2, 
\end{equation}
where $\tau(\eta) = 2 r(\eta)/c$, assuming $r(\eta) = \|\vec{p}^{\,\text{SAR}}(\eta) - \vec{p} \|$ is the distance between the corner reflector and the \ac{SAR} platform.  The \ac{SAR} platform follows a straight path, perpendicular to its line of sight at a distance $r_0$, as shown in \Cref{fig:aspect_angle} and \Cref{fig:azimuth_angle}. Its position along the path is given by:
\begin{equation}
    \vec{p}^{\,\text{SAR}}(\eta) = \begin{pmatrix}
        \cos(\phi^a) & -\sin(\phi^a) & 0 \\
        \sin(\phi^a) & \cos(\phi^a) & 0 \\
        0 & 0 & 1
    \end{pmatrix}
    \begin{pmatrix}
        r_0 \sin(\theta^a) \\
        \eta v \\
        r_0 \cos(\theta^a)
    \end{pmatrix}.
\end{equation}

To evaluate the far-field integrals $N^{\theta}_p$ and $N^{\phi}_p$, we assume a narrowband waveform, allowing the wavenumber inside the surface integrals to be approximated as constant at center frequency, $k \approx 2\pi f_0/c$, where $c$ is the speed of light~\cite{polycarpou1995radar}:
\begin{equation}\label{Eq:far_field_integrals}
\begin{aligned}
    N^{\theta}_p &= \iint_{\mathcal{A}^i_p} \left( \vec{J}_p \cdot
    \begin{pmatrix}
        \cos(\theta')\cos(\phi') \\
        \cos(\theta')\sin(\phi')\\
         -\sin(\theta') 
    \end{pmatrix}
    \right) e^{j k L} \dA, \\
    N^{\phi}_p &= \iint_{\mathcal{A}^i_p} \left( \vec{J}_{p} \cdot
    \begin{pmatrix}
        -\sin(\phi') \\
        \cos(\phi') \\
        0
    \end{pmatrix}\right) e^{j k L} \dA, \\
    L &= x \sin(\theta')\cos(\phi') + y  \sin(\theta')\sin(\phi') + z \cos(\theta'),
\end{aligned}
\end{equation}
where $\mathcal{A}^i_p$ is the illuminated area on the final reflecting plate, and $\vec{J}_{p}$ is the current density on the final reflecting plate $\mathcal{A}_s$, $s\in\{1,2,3\}$ of reflection path $p$:
\begin{equation}\label{Eq:currentDensity}
    \vec{J}_{p} = 2 \hat{n}_s \times \vec{H}_{p} = 2 \hat{n}_s \times \frac{A^t}{Z_0} e^{-j k (\hat{k}_{p}\cdot \vec{r})} \hat{h}_{p},
\end{equation}
where $\vec{H}_{p}$ is the incident magnetic field, $\hat{n}_s$ is the normal vector of plate $\mathcal{A}_s$, $Z_0$ is the intrinsic impedance of free space, and $\vec{r} = [x, y, z]^T$, $\hat{k}_{p}$ and $\hat{h}_{p}$ denote the direction of travel and polarization of $\vec{H}_p$ respectively.

Thus, the far-field integrals $N^{\theta}_p$, $N^{\phi}_p$ produced by each reflection path $p$ depend on the incident magnetic wave $H_p$ and the illuminated area of the final reflecting plate $\mathcal{A}^i_p$. For single-bounce reflection paths, the reflecting plate is entirely illuminated, i.e., $\forall p \in [1, 2, 3]:\mathcal{A}^i_p = \mathcal{A}_s$ for $\theta', \phi' \in \left[0, \pi/2\right]$; and the incident magnetic field is identical to the transmitted field, i.e., $\forall p \in [1, 2, 3]: H_p = H^t$, defined by direction of travel $\hat{k}^t$ and polarization $\hat{h}^t$:
\begin{equation}
    \hat{k}^t = -\begin{pmatrix}
        \sin(\theta') \cos(\phi')\\
        \sin(\theta') \sin(\phi')\\
        \cos(\theta')
    \end{pmatrix}, \;
    \hat{h}^t = \begin{pmatrix}
        \cos(\theta')\cos(\phi') \\
        \cos(\theta')\sin(\phi') \\
        -\sin(\theta')
    \end{pmatrix}.
\end{equation} 

For multi-bounce scattering (e.g., double- and triple-bounce paths), intermediate reflections are modeled using \ac{GO}. where each reflection is treated as specular. For a path $p$, sequential specular projection determines the incident wave vector $\hat{k}_p$, magnetic polarization $\hat{h}_p$, and the constrained illuminated domain $\mathcal{A}^i_p$ on the final surface. The corresponding far-field \ac{PO} integrals are detailed in ~\ref{appendix:components}.

\subsection{Backscatter measurement and image formation}\label{sec:measure}

\Ac{SAR} systems typically record a single polarization channel. In this work, the system is assumed to operate in HH mode, so measurements are performed in horizontal polarization, denoted by $\mathcal{M}^H$, corresponding to measuring $\hat{\phi}$ component:
\begin{equation}
\begin{aligned}
    \mathcal{M}^H\left(E^r(t, \eta)\right) &= \Re \left\{ A^r(\eta) \rect\left(\frac{t - \tau(\eta)}{T}\right) e^{j \psi(t, \eta)} \right\}, \\
    A^r(\eta) &= \frac{-j Z_0}{2 \lambda r(\eta)} \sum_{p \in \mathcal{P}} N^{\phi}_p. \\
\end{aligned}
\end{equation}
Conversely, measuring the vertical polarization would correspond to measuring the $\hat{\theta}$ component. Afterwards, the measured signal is demodulated using \ac{QD}, by mixing the received signal with the complex carrier signal $e^{-j2\pi f_0 t}$, and applying a \ac{LPF} to isolate the baseband component \cite{jansing2021introduction}:
\begin{equation}
\begin{aligned}
     \text{QD}\left( \mathcal{M}^H\left(E^r(t, \eta) \right) \right) =&\; \text{LPF}\left(e^{-j2\pi f_0 t} \mathcal{M}^H \left( E^r(t, \eta) \right)\right), \\
     \text{QD}\left( \mathcal{M}^H\left( E^r(t, \eta)\right) \right) =&\; \frac{A^r(\eta)}{2} \rect\left(\frac{t - \tau(\eta)}{T}\right) \\
     & \hspace{4em} e^{j\pi\sbr{-2f_0\tau(\eta) + K(t - \tau(\eta))^2}}.
\end{aligned}
\end{equation}

Subsequently, \ac{RDA} \cite{raney1994precision} is applied to focus the complex demodulated signal into a complex-valued \ac{SAR} image. Magnitude detection is then performed to obtain the real-valued \ac{SAR} image.

\subsection{Computational complexity}

As evident from \cref{eq:loss}, each evaluation of the loss function requires computing the \ac{SAR} image generated by the physical perturbation $\mathcal{O}(\tilde{\mathcal{S}}(\Theta), \theta^a, \phi^a)$ and superposing it with the clean image, across $N$ samples. The superposition step is negligible compared to computing the \ac{SAR} image, which has two computationally expensive components: \Circled{1} the evaluation of the integrals within \cref{Eq:spherical_components} required to obtain the far-field components, and \Circled{2} the \acp{FFT} and \acp{IFFT} required within the \ac{RDA}.

The far-field integrals are analytically intractable, and must therefore be computed numerically. Assuming each integral is approximated using $P$ samples, this step has complexity $O(P)$. If the size and shape of the corner reflector are known in advance, the far-field components can be precomputed and stored in a lookup table for all $\theta', \phi' \in [0, \pi/2]$, eliminating this cost during optimization.

The \ac{RDA} requires an \ac{FFT} and \ac{IFFT} in fast and slow time. Therefore, assuming $F$ fast time samples, and $S$ slow time samples, the computational complexity to focus one \ac{SAR} image using \ac{RDA} is $O(SF \log(FS)).$ Thus, assuming cached far-field components, the computational complexity of a single loss evaluation is $O(N FS \log(FS))$.

\section{Dataset}\label{sec:dataset}

Short of evaluating the proposed attack on a \ac{SAR}, a dataset needs to be generated with a simulator or acquired from a third party. In the absence of a simulator capable of simulating a wide mix of material properties, the \ac{MSTAR} dataset \cite{ross1998standard} is chosen. Developed in the 1990s by \ac{DARPA} and \ac{AFRL}, the \ac{MSTAR} dataset remains widely used, including for the evaluation of \acp{PAA} \cite{peng2022scattering, xia2023sar-pega, luo2024sar-patt, xie2024migaa, zhang2024physically, ma2025sar-paa}.

The \ac{MSTAR} dataset contains labeled high-resolution X-band \ac{SAR} images of military vehicles and targets, specifically 2S1, BMP-2, BRDM-2, BTR-60, BTR-70, D7, T-62, T-72, ZIL-131, and ZSU-234. The HH-polarized samples were captured in spotlight mode, quadratically-demodulated and focused using the \ac{RDA}, consistent with the assumptions made in \Cref{sec:assumptions_sar_system}. The data was acquired at four different incidence angles. For each incidence angle, the full $[0, 2\pi]$ azimuthal range is densely populated with samples. For a given class, all samples are images of the same physical scene, captured from different aspect angles. A detailed summary of the incidence angles and the number of samples per class per angle is provided in \Cref{tab:mstar_samples}. The high variability of aspect angles makes the \ac{MSTAR} dataset well suited for evaluating \ac{SAIPAS}.  Other SAR datasets are available \cite{wang2023category, zhang2021sar, zhang2020ls-ssdd, huang2018opensarship}, however, they lack repeated observations of the same scene across various aspect angles, making them unsuitable for evaluating \ac{SAIPAS}.

\begin{table}[ht]
    \caption{\label{tab:mstar_samples}Number of samples per class in the \ac{MSTAR} dataset, classified by the incidence aspect angle $\theta^a$ (degrees).}
    \centering
    \begin{tabularx}{\linewidth} {
        >{\raggedright\arraybackslash}X 
        >{\centering\arraybackslash}X
        >{\centering\arraybackslash}X 
        >{\centering\arraybackslash}X
        >{\centering\arraybackslash}X}
        \toprule
        Class label & $75^{\circ}$ & $73^{\circ}$ & $60^{\circ}$ & $45^{\circ}$ \\
        \midrule
        2S1 & 274 & 299 & 288 & 303 \\
        BMP-2 & 195& 233 & 0 & 0 \\
        BTR-60 & 195 & 256 & 0 & 0 \\
        BTR-70 & 196 & 233 & 0 & 0 \\
        D7 & 274 & 299 & 0 & 0 \\
        T-62 & 273 & 299 & 0 & 0 \\
        T-72 & 196 & 232 & 0 & 0 \\
        ZIL-131 & 274 & 299 & 0 & 0 \\
        ZSU-234 & 274 & 299 & 406 & 422 \\
        BRDM-2 & 274 & 298 & 420 & 423\\
        \bottomrule
    \end{tabularx}
\end{table}

\subsection{Bounding boxes}\label{sec:bb}

Each \ac{MSTAR} sample contains a single target roughly centered in the image, although small misalignments occur across samples. Applying a static pixel-space perturbation fails to account for these offsets, causing physical discrepancy between the perturbation and the target's physical coordinate system. By constructing bounding boxes around the target, a consistent spatial reference is provided, allowing the perturbation to be correctly mapped to the scene and compensating for misalignments across samples. Simple methods for defining bounding boxes, such as selecting the brightest rectangle, are unreliable, because only the portion of the target facing the radar produces a strong return. While alternative approaches exist \cite{lin2023sived, yang2023yang}, a method tailored to the densely sampled, aspect-angle–annotated \ac{MSTAR} dataset was developed for improved simplicity and reliability.

\begin{figure}[ht]
    \centering
    \subfloat[]{\includegraphics[width=0.49\linewidth]{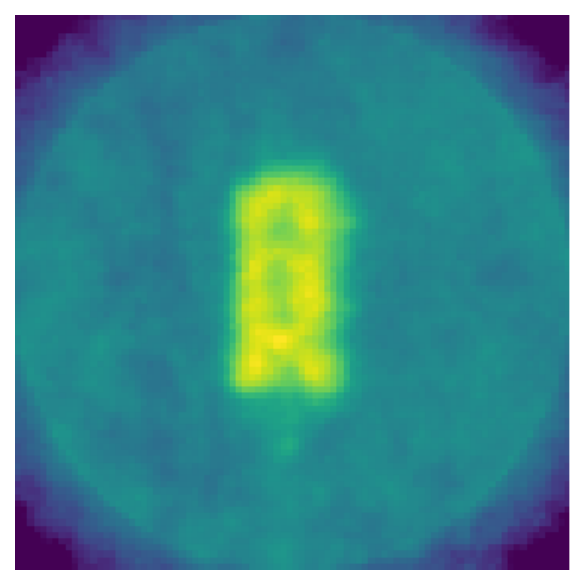}\label{fig:composite_no_bounding_box}}
    \hfil
    \subfloat[]{\includegraphics[width=0.49\linewidth]{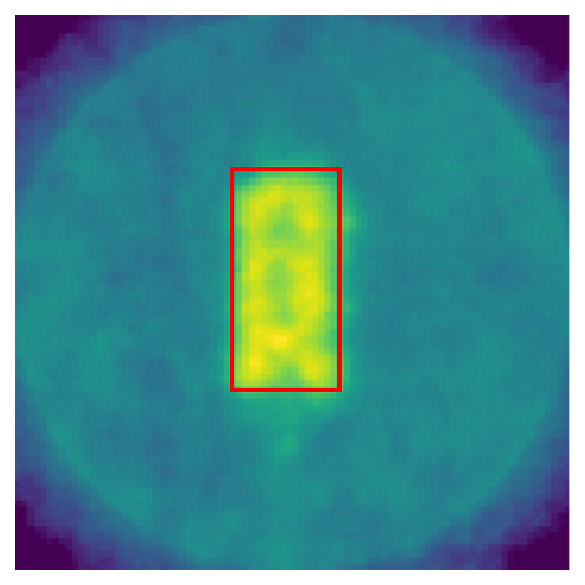}\label{fig:composite_bounding_box}}
    \caption{Composite image created by aligning and averaging observations of the T-62 for $\theta^a = 75^\circ$: (a) without bounding box and (b) with bounding box $R^{\text{ref}}$.}
\end{figure}

The proposed bounding-box method begins with the estimation of the area occupied by a target object. The dense azimuth sampling of \ac{MSTAR} allows an object's area to be inferred across samples from multiple azimuth angles. To establish the static physical dimensions of a target class for a given scene and incidence angle $\theta^a$, all image samples are rotated by $-\phi^a$ to align their aspect angles and then averaged into a composite image, as shown in \Cref{fig:composite_no_bounding_box}. The composite image is converted to a logarithmic scale and normalized. A bounding box is then constructed by $\delta$-thresholding the image to create a binary mask and fitting a minimum-area rotated rectangle $R^{\text{ref}}$ (also known as oriented bounding box), around the largest contour, as shown in \Cref{fig:composite_bounding_box}.

For each image of the same scene and incidence angle $\theta^a$, a rectangle $R$ with the same dimensions as $R^{\text{ref}}$ is positioned with its rotation aligned to the sample's azimuth aspect angle $\phi^a$. Prior to fitting, the images are preprocessed, specifically logarithmically scaled, thresholded at 0.5 to suppress background clutter, and gamma-corrected at 1.5. To localize $R$, a loss function is defined, consisting of two components: \Circled{1} the pixel-based loss $\mathcal{L}_{\text{pixel}}$ and \Circled{2} the distance-based loss $\mathcal{L}_{\text{dist}}$.

The pixel-based loss $\mathcal{L}_{\text{pixel}}(R)$: This component rewards bright pixels near the bottom of the bounding box, corresponding to the side of the target facing the \ac{SAR} system. This reflects the inherent property of \ac{SAR} images, in which surfaces oriented toward the radar produce stronger returns:
\begin{equation}
    \mathcal{L}_{\text{pixel}}(R) = - \frac{\sum_{x, y} I(x, y) \mathcal{M}\left(x, y, R\right) d(y, R)^\alpha}{|R|},
\end{equation}
where $|R|$ is the number of pixels in the rectangle, $\alpha$ is a hyperparameter, $I(x, y)$ is the intensity of the processed image at pixel $(x, y)$, $\mathcal{M}(x, y, R)$ is the binary mask of $R$, $d(y, R)$ is a vertical weighting:
\begin{equation}
    d(y, R) = \frac{y - y^R_{\text{min}}}{y^R_{\text{max}} - y^R_{\text{min}}}.
\end{equation}
    
The distance-based loss $\mathcal{L}_{\text{dist}}(x^R, y^R)$: This component penalizes displacement from the center of $R^{\text{ref}}$ in the composite image, denoted by $(x^\text{ref}$, $y^\text{ref})$:
\begin{equation}
    \mathcal{L}_{\text{dist}}(x^R, y^R) = \left(\frac{(x^R - x^\text{ref})^2 + (y^R - y^\text{ref})^2}{d_{\text{max}}}\right)^\beta,
\end{equation}
where $d_{\text{max}} = \sqrt{x_{\text{max}}^2 + y_{\text{max}}^2}$ is the maximum displacement allowed, and $\beta$ is a hyperparameter.

\begin{figure}[ht]
    \centering
    \subfloat[]{\includegraphics[width=0.46\linewidth]{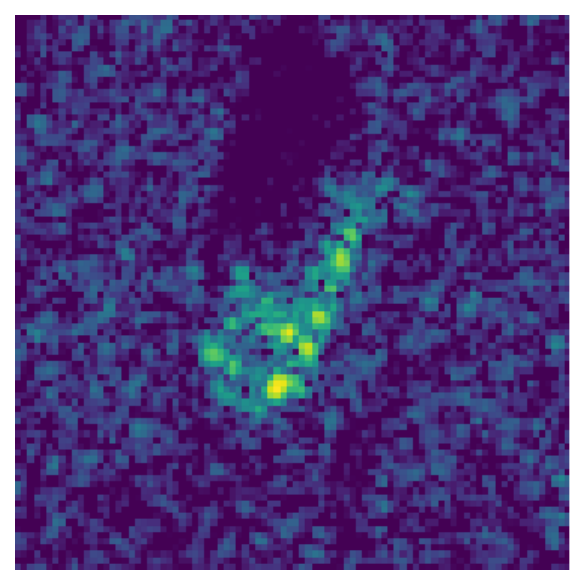}\label{fig:boxes_example_no_box}}
    \hfil
    \subfloat[]{\includegraphics[width=0.53\linewidth]{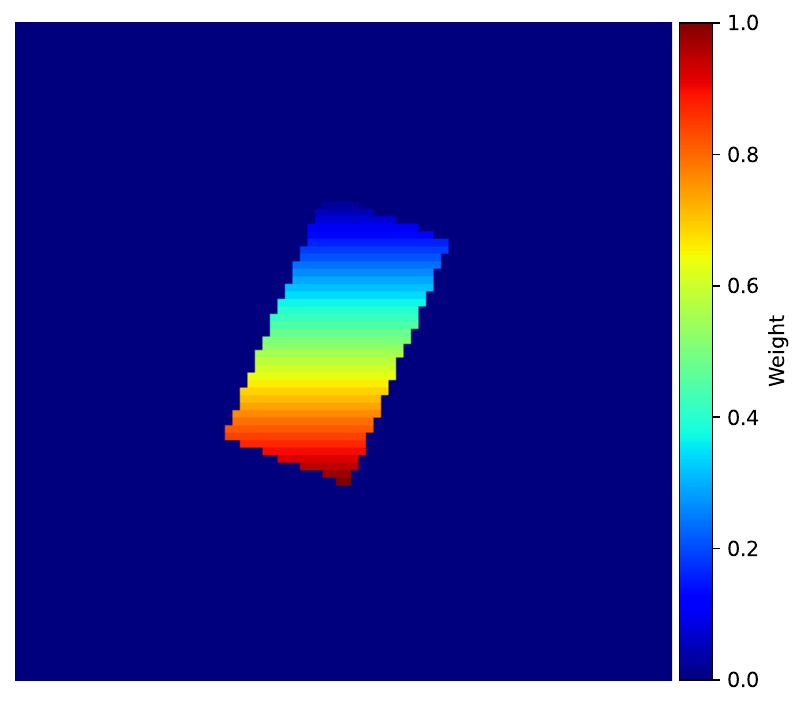}\label{fig:boxes_example_mask}}
    \\
    \subfloat[]{\includegraphics[width=0.46\linewidth]{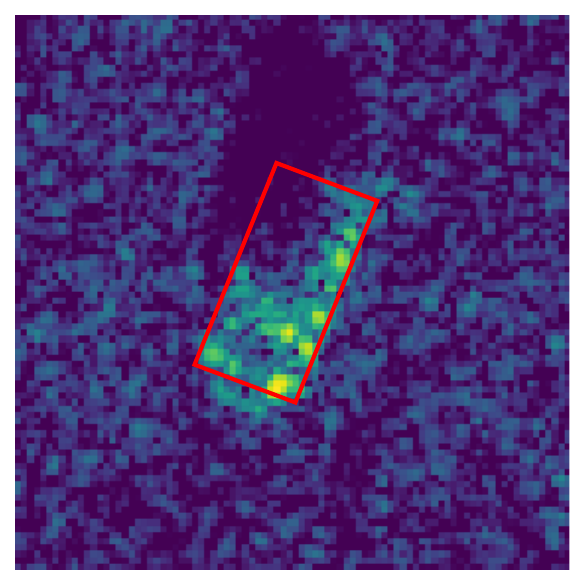}\label{fig:boxes_example_box}}
    \caption{A \ac{SAR} image sample of (a) a T-62 observed from $\theta^a = 75^\circ$ and $\phi^a = 200.2^\circ$, (b) the weighted mask $\mathcal{M}\left(x, y, R\right) d(y, R)^\alpha$ used to find $R$, and (c) the bounding box $R$ found by solving \eqref{eq:bb}.}
\end{figure}

The bounding box is obtained by solving:
\begin{equation}\begin{split}\label{eq:bb}
    \min_{x^R, y^R}\;& \mathcal{L}_{\text{pixel}}(R) + \lambda \mathcal{L}_{\text{dist}}(x^R, y^R), \\
    \text{s.t.}\;& x^R \in [-x_{\text{max}}, x_{\text{max}}] \land y^R \in [-y_{\text{max}}, y_{\text{max}}],
\end{split}\end{equation}
where $\lambda$ is a hyperparameter controlling the relative weight of the distance-based loss. Because MSTAR lacks ground-truth spatial bounding box annotations, the hyperparameters ($\delta = 0.7, \alpha = 1.5, \beta = 0.5, \lambda = 0.1$) were selected via grid search and qualitatively verified. As a demonstration, \Cref{fig:boxes_example_box} shows the bounding box for the sample in \Cref{fig:boxes_example_no_box}, utilizing the weighted mask $\mathcal{M}\left(x, y, R\right) d(y, R)^\alpha$ shown in \Cref{fig:boxes_example_mask}.

\subsection{Data sampling for evaluating \ac{SAIPAS}}

\ac{SAIPAS} optimizes an adversarial perturbation for each class in the \ac{MSTAR} dataset. To achieve this, for a given target class, we construct a training subset $\mathcal{G} \subset \mathcal{D}$ by sampling images at discrete nominal azimuth intervals $\Delta\phi^a$ for each available incidence angle. Because \ac{MSTAR} features discrete and slightly non-uniform azimuth sampling, exact grid points at multiples of $\Delta\phi^a$ are rarely present. To accommodate this discretization, we introduce an azimuth tolerance $\epsilon$, which defines an acceptance interval $[\phi - \epsilon, \phi + \epsilon]$ around each nominal target angle $\phi$. \Cref{alg:subset_sampling} summarizes our sampling strategy. 

For each perturbation optimization, a training set is derived from the dataset, with a default azimuth spacing of 10$^\circ$ and a tolerance of 5$^\circ$.  The training set is sampled coarsely to limit the computational cost of training. The remainder of the dataset is used as a validation set.

To thoroughly evaluate aspect-independent attack effectiveness across the full aspect angle domain, training and test samples are drawn from the full \ac{MSTAR} dataset. To maintain dense angular sampling required for physical radar simulation, samples from the\ac{ATR} training set are included. 

\begin{algorithm}[ht]
\caption{Sampling images for a given class and azimuth spacing}\label{alg:subset_sampling}
\begin{algorithmic}[1]
\Require Set of available samples $\mathcal{D}$ for class label $l$, azimuth spacing $\Delta\phi^a$, tolerance $\epsilon$
\Ensure Subset of samples $\mathcal{G}$ uniformly distributed in azimuth for each incidence angle
\State $\mathcal{G} \leftarrow \emptyset$ \Comment{Initialize empty set of selected samples}
\For{each incidence angle $\theta$ present in $\mathcal{D}$}
    \State $\phi_0^a \sim U(0, \Delta\phi^a)$ \Comment{Randomize azimuth starting point}
    \State $\phi \leftarrow \phi_0^a$
    \While{$\phi < 2\pi$}
        \State Select $s$ such that $\theta^a_s = \theta \wedge |\phi^a_s - \phi| \leq \epsilon$
        \State $\mathcal{G} \leftarrow \mathcal{G} \cup \{s\}$
        \State $\phi \leftarrow \phi + \Delta\phi^a$
    \EndWhile
\EndFor
\State \Return $\mathcal{G}$ 
\end{algorithmic}
\end{algorithm}

\section{Results and discussion}\label{sec:exp}

This section presents the experimental evaluation of the proposed \ac{SAIPAS}. Different optimization strategies were compared. Comprehensive experiments were conducted to evaluate the attack performance of \ac{SAIPAS} under various conditions, such as different numbers of reflectors, limited training data, and transferability to unseen models. Additionally, \ac{SAIPAS} was evaluated under the assumption that the attacker has partial knowledge of the aspect angles.

\subsection{Experimental setup}\label{sec:exp_setup}

This section describes the experimental setup used to evaluate \ac{SAIPAS}. We first specify the \ac{SAR} system parameters and evaluation metrics, then introduce the \ac{ATR} models used for assessment, and finally outline the candidate optimization algorithms considered for solving \Cref{eq:loss}.

\subsubsection{Adversarial reflectors}
The adversarial reflectors are trihedral corner reflectors of dimensions $0.3\text{m} \times 0.3\text{m} \times 0.3\text{m}$ (comparable to the $0.4\text{m} \times 0.4\text{m} \times 0.2\text{m}$ reflectors reported in prior work \cite{ma2025sar-paa}). Unless otherwise stated, four corner reflectors are used per configuration, as this is the minimal number of reflectors required to allow full azimuth coverage.

\subsubsection{SAR system specification}

The reflected signal $E^r(t,\eta)$ and image formation depend on the \ac{SAR} system's technical specifications. Certain parameters, such as range $r$, center frequency $f_0$, bandwidth $B = K T$, pixel ground spacing, and polarization can be extracted from the provided metadata \cite{ross1998standard}. The remaining parameters, including pulse duration $T$, platform speed $v$, pulse repetition frequency, and sample rate were estimated based on typical values reported for comparable \ac{SAR} platforms \cite{perna2019imaging, walls2014multi, horn2017fsar, Giovanni2004current}. Following Quadrature Demodulation and low-pass filtering detailed in \Cref{sec:measure}, the complex baseband signal sampling rate $f_s$ was set to $600\text{ MHz}$ to satisfy the Nyquist criterion for complex analytic signals ($f_s > B$)~\cite{shannon1949communication}.  \Cref{tab:simulated_sar_specifications} specifies the values of all system parameters in use. Sidelobe suppression during Range-Doppler processing is achieved by applying Taylor windowing with a target peak sidelobe level of $-30\text{ dB}$ and transition order $\bar{n} = 4$ across both range and azimuth domains.

\begin{table}[ht]
    \caption{Specification of the simulated \ac{SAR} system.}
    \label{tab:simulated_sar_specifications}
    \centering
    \begin{tabularx}{\linewidth}{
        >{\raggedright\arraybackslash}X 
        >{\centering\arraybackslash}X
        }
        \toprule
        Variable & Value \\
        \midrule
        Range & \{4500 m , 5000 m\} \\
        Platform speed & 50 m/s \\
        Center frequency  &  9.6 GHz  \\ 
        Bandwidth  & 591 MHz \\
        Pulse duration  & 5 $\mu$s \\
        Sample rate & 600 MHz \\
        Pulse repetition rate  & 1200 Hz \\ 
        Ground sample distance & $0.3\text{ m} \times 0.3\text{ m}$ \\
        Polarization & HH \\
        \bottomrule
    \end{tabularx}
\end{table}

The absolute amplitude of the transmitted signal, $A^t$, cannot be directly determined from the \ac{MSTAR} metadata, as the images are stored in a relative, arbitrary scale. To account for this unknown scaling, the amplitude of the transmitted pulse $A^t$ was chosen so that a corner reflector of dimensions $0.3\text{ m} \times 0.3\text{ m} \times 0.3\text{ m}$, observed from its boresight, yields a peak intensity equal to half the average maximum pixel intensity across all target classes at their respective brightest aspect angles. This ensures that the simulated reflector returns operate within the actual dynamic range of strong target scatterers. Under this configuration, the amplitude of the return signal (i.e. perturbation strength) may be varied by changing the physical dimensions of the corner reflector. 

\subsubsection{Evaluation metrics}

Fooling rate \cite{moosavi-dezfooli2017universal} or attack success rate \cite{zhang2022investigating} is originally called error rate \cite{goodfellow2015explaining}. The fooling rate of an attack $\mathcal{A}$ against a model $f$ applied to dataset $\mathcal{D}$, is the proportion of data that is correctly classified by $f$ in the absence of $\mathcal{A}$, but is misclassified by $f$ in the presence of $\mathcal{A}$:
\begin{equation}\label{eq:foolingrate:ori}
    \frac{\sum_{\bX\in\mathcal{D}}\mathbb{I}\cbr{\hat{y}(\bX) = l \wedge \hat{y}(\mathcal{A}(\bX)) \neq l}}{\sum_{\bX\in\mathcal{D}}\mathbb{I}\cbr{\hat{y}(\bX) = l}},
\end{equation}
where $\mathbb{I}$ is the indicator function, $\hat{y}(\bX)$ is the predicted label for image $\bX$, and $l$ the true label of $\bX$.

For each scene, the fooling rate is computed using Eq.~\eqref{eq:foolingrate:ori}. The fooling rates for all classes are then averaged to produce the \emph{average fooling rate}, which is the main metric for evaluating attack efficacy in this study.

\subsubsection{\ac{ATR} models}\label{sec:atr_models}

\ac{SAIPAS} was evaluated against AConvNet \cite{chen2016target}, a model specifically designed for \ac{SAR} \ac{ATR} with a strong performance \cite{chen2016target}. To assess transferability across architectures, four additional widely used convolutional networks were implemented: AlexNet \cite{krizhevsky2012imagenet}, DenseNet-121 \cite{huang2017densely}, MobileNetV2 \cite{sandler2018mobilenetv2}, and ResNet50 \cite{he2016deep}. These models were chosen to represent a diverse set of architectures, and their extensive usage in prior adversarial machine learning studies targeting \ac{SAR} \ac{ATR} \cite{peng2022scattering, xie2024migaa, zhang2024physically, ma2025sar-paa}. 

All models were trained with standard \ac{SAR} data augmentations (sliding-window translation, random rotation, scaling, and additive random noise) to emulate realistic \ac{SAR} imaging variability and prevent overfitting \cite{ding2016convolutional, furukawa2017deep}. The model weights were trained using a standard 70/20/10 train/validation/test split on the \ac{MSTAR} dataset. Training was performed using \ac{SGD}, with a batch size of 32, a learning rate of 0.001, and a momentum of 0.9 for 100 epochs. \Cref{tab:models} shows the test accuracies.

\begin{table}[ht]
    \centering
    \caption{A summary of target \ac{ATR} models.}
    \begin{tabularx}{\linewidth} {
        >{\raggedright\arraybackslash}X 
        >{\centering\arraybackslash}X
    }
    \toprule
        Model & Test accuracy \\
    \midrule
        AConvNet & 99.6\% \\
        AlexNet & 99.7\% \\
        DenseNet-121 & 99.3\% \\
        MobileNetV2 & 98.8\% \\
        ResNet50 & 99.2\% \\
    \bottomrule
    \end{tabularx}
    \label{tab:models}
\end{table}

\subsubsection{Optimization algorithms}

Four optimization algorithms were investigated. Three evolutionary algorithms, specifically \ac{DE} \cite{storn1997differential} and \ac{PSO} \cite{kennedy1995particle}, and \ac{RCGA} \cite{deb1995simulated} were selected based on their effectiveness in adversarial optimization tasks against \ac{SAR} \ac{ATR} models \cite{xie2024migaa, zhang2024physically, ma2025sar-paa} and their ability to navigate non-convex, discontinuous loss surfaces. For diversity, \ac{BO} \cite{snoek2012practical} was included as a model-based alternative, motivated by its potential for sample-efficient search given the low dimensionality of the search space (e.g., $13$ variables for $4$ reflectors).

\subsection{Fine-tuning optimization}\label{sec:finetuning}

The first set of experiments focused on fine-tuning the optimization procedure for maximizing the average fooling rate. Specifically, optimization algorithms were compared, and for the top-performing optimization algorithm, the impact of hyperparameter variation on optimization performance was studied. The impact of the choice of optimization variables (which angles of the corner reflectors to fix, and which angles to optimize) on attack performance was also studied.
These experiments established the baseline configuration adopted in the remainder of this work.

\begin{table}[th]
    \caption{Summary of hyperparameters and average fooling rates for various optimizers.}
    \label{tab:optimizers}
    \centering
    \renewcommand{\arraystretch}{1.3} 
    \begin{tabularx}{\linewidth}{
        >{\centering\arraybackslash}m{0.1\linewidth}  
        >{\centering\arraybackslash}X  
        >{\centering\arraybackslash}m{0.16\linewidth}  
    }
    \toprule
    Optimizer & Hyperparameters & Avg. fooling rate \\
    \midrule
    \ac{BO} & 
    \makecell{
        Nr. of initial points = 150 \\ 
        Max iterations = 2250\\ 
        Smoothness of kernel = 2.5 \\ 
        Exploration parameter = 0.1
    } & 35.01\% \\
    \midrule 
    \makecell{\ac{DE}} & 
    \makecell{
        Population size = 40 \\ 
        Max iterations = 60\\ 
        Mutation factor = 0.5\\ 
        crossover probability = 0.3\\ 
    } & 48.60\% \\
    \midrule 
    \ac{PSO} & 
    \makecell{
        Nr. of particles = 40 \\ 
        Max iterations = 60\\ 
        Cognitive learning rate = 0.6\\ 
        Social learning rate = 1.0\\ 
        Inertia weight = 0.8
    } & 48.09\% \\
    \midrule
    \makecell{\ac{RCGA}} & 
    \makecell{
        Population size = 40 \\ 
        Max iterations = 60\\ 
        Mutation = 0.5\\ 
        Recombination = 0.7\\ 
        Crowding for crossover = 20\\ 
        Mutation probability = 0.2\\ 
        Tournament selection size = 3
    } & \makecell{\textbf{63.6\%}} \\
    \bottomrule
    \end{tabularx}
\end{table}

\subsubsection{Varying optimization algorithm}

\begin{figure}[th]
    \centering
    \includegraphics[width=0.95\linewidth]{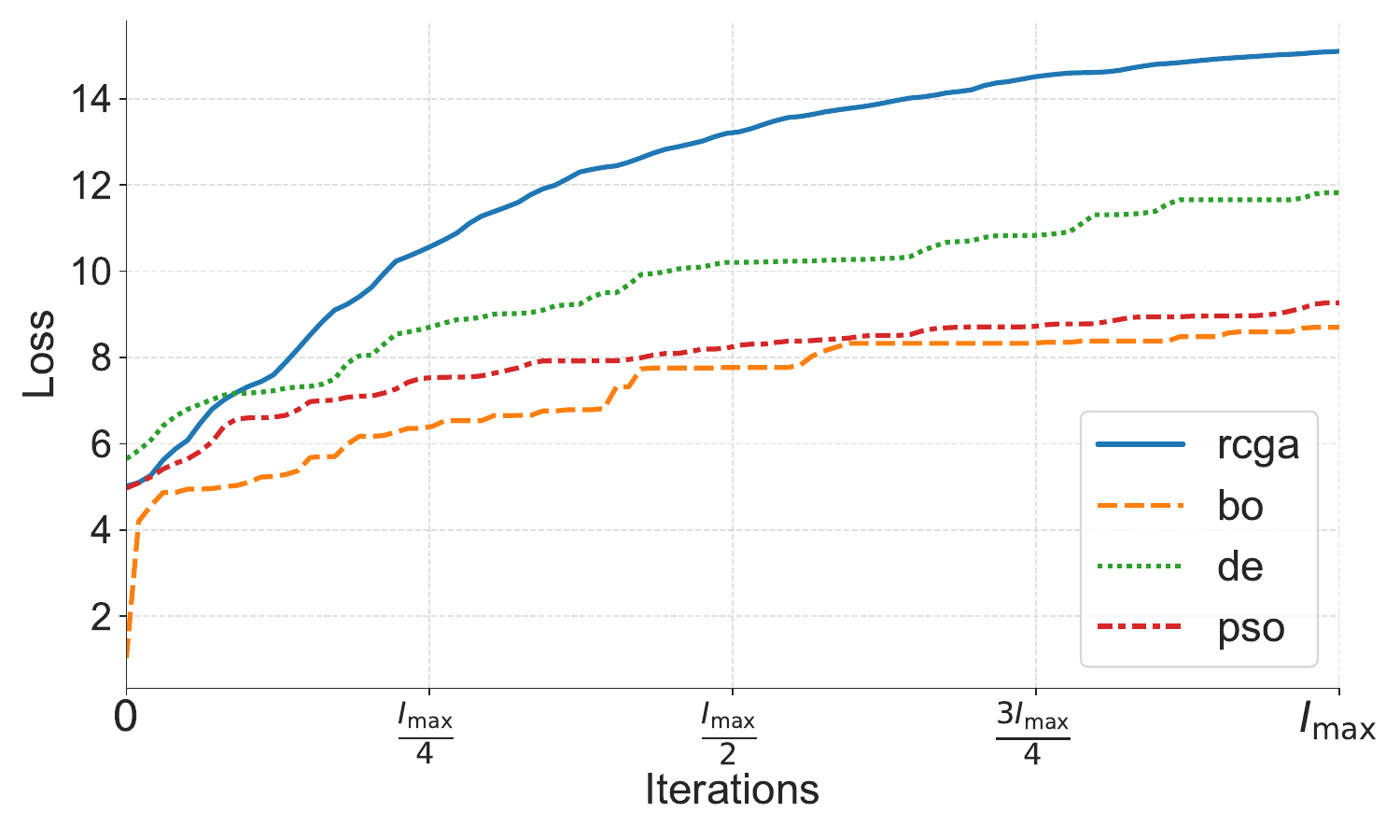}
    \caption{Average loss per iteration for each optimizer, where $I_{\text{max}}$ is the maximum number of iterations.}
    \label{fig:experiment_1}
\end{figure}

\Ac{PSO}, \ac{DE}, \ac{RCGA}, and \ac{BO} were investigated. The number of loss function evaluations was held constant across all algorithms. \Cref{tab:optimizers} summarizes the configuration of each optimizer and the corresponding average fooling rates. \Ac{RCGA} achieved the highest fooling rate and final loss, as shown in \Cref{fig:experiment_1}. \Ac{BO} significantly underperformed, potentially due to the unsatisfactory fit of a Gaussian process to the objective function.

\subsubsection{Varying optimizer hyperparameters}

\begin{figure}[ht]
    \centering
    \includegraphics[width=0.95\linewidth]{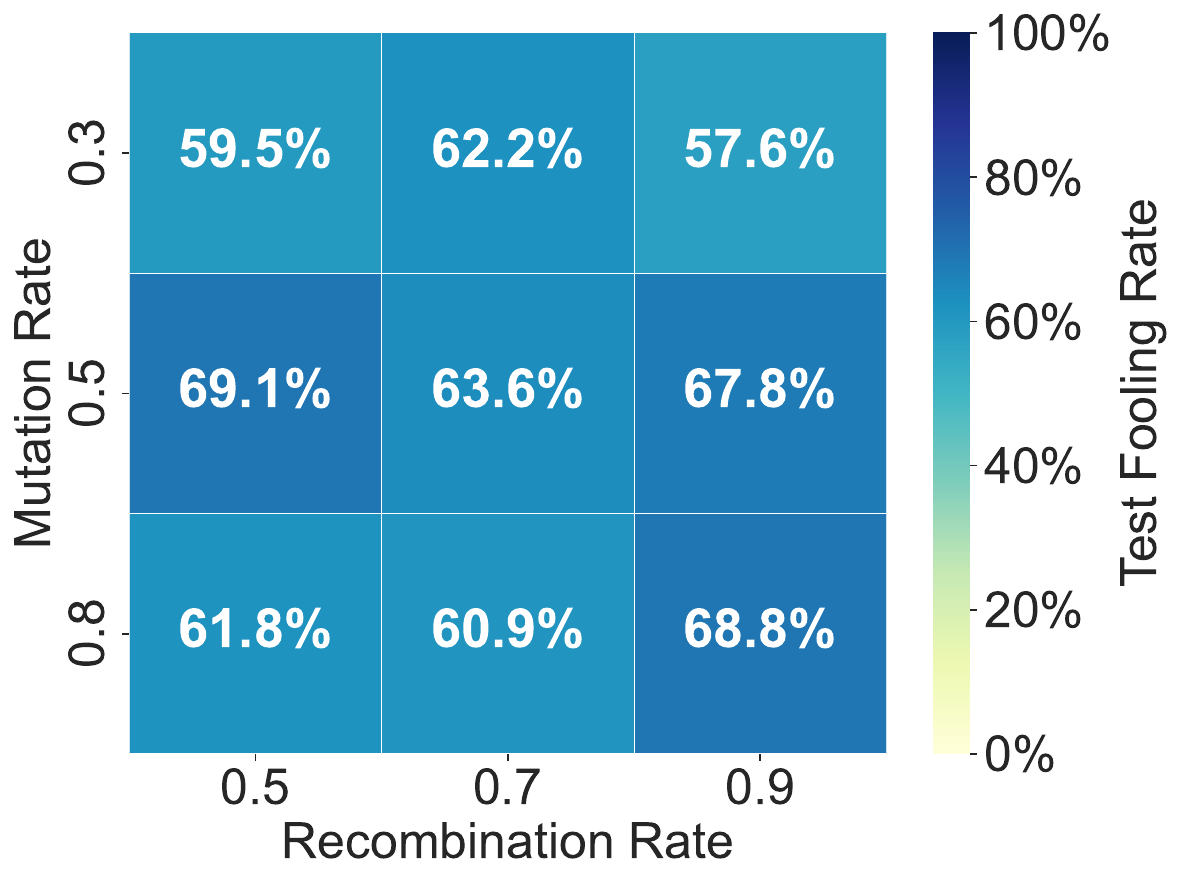}
    \caption{Average Fooling rates for each mutation and recombination rate for \ac{RCGA}.}
    \label{fig:experiment_2}
\end{figure}

The \ac{RCGA} optimizer was further finetuned with a hyperparameter study. The mutation rate was varied between $\{0.3, 0.5, 0.8\}$, and the recombination rate was varied between $\{ 0.5, 0.7, 0.9 \}$, while other hyperparameters were kept constant. \Cref{fig:experiment_2} summarizes the average fooling rates for each configuration The configuration with 0.5 mutation rate and 0.5 recombination rate achieved the best average fooling rate. 

\subsubsection{Varying optimization variables}

Each corner reflector is parameterized by its boresight incidence angle, $\theta$, and its boresight azimuth angle, $\phi$. We evaluate whether reducing search space dimensionality improves attack performance by fixing these parameters. Setting $\theta_i = \arctan\sqrt{2}$ for all $i \in \{1, 2, 3, 4\}$ and orthogonal azimuths $\phi_k = \frac{k\pi}{2}$ for $k \in \{0, 1, 2, 3\}$ ensures full angular coverage.

\begin{table}[ht]
    \caption{Results for different choice of optimization variables.}
    \label{tab:fix_orientation}
    \centering
        \begin{tabularx}{\linewidth}{
        >{\centering\arraybackslash}c
        >{\centering\arraybackslash}c 
        >{\centering\arraybackslash}c
        >{\centering\arraybackslash}X
        }
        \toprule
        Configuration & Fixed $\theta$'s & Fixed $\phi$'s & Average fooling rate \\
        \midrule
        1 & False & False & \textbf{69.1}\%\\
        2 & True & False & 64.8\%\\
        3 & False & True & 60.0\%\\
        4 & True & True & 59.8\%\\
        \bottomrule
        \end{tabularx}
\end{table}

As \Cref{tab:fix_orientation} shows, fixing the incidence angle $\theta$ and $\phi$ degrades performance. This indicates fine-tuning backscatter amplitude, which is governed by these orientation angles, is vital for maximizing fooling efficacy.

Concluding this subsection on optimization fine-tuning, using \ac{RCGA} with 0.5 mutation rate and 0.5 recombination rate, while allowing the reflector's orientation to be optimized yields the best attack performance, achieving an average fooling rate of $69.1\%$. Consequently, all subsequent experiments use this configuration.

\subsection{Evaluating attack performance}\label{sec:eval_attack}

Experiments were performed to assess the average fooling rate under various conditions.

\subsubsection{Visualizations}

\begin{figure*}[!t]
    \centering
    \subfloat[]{\includegraphics[width=0.49\linewidth]{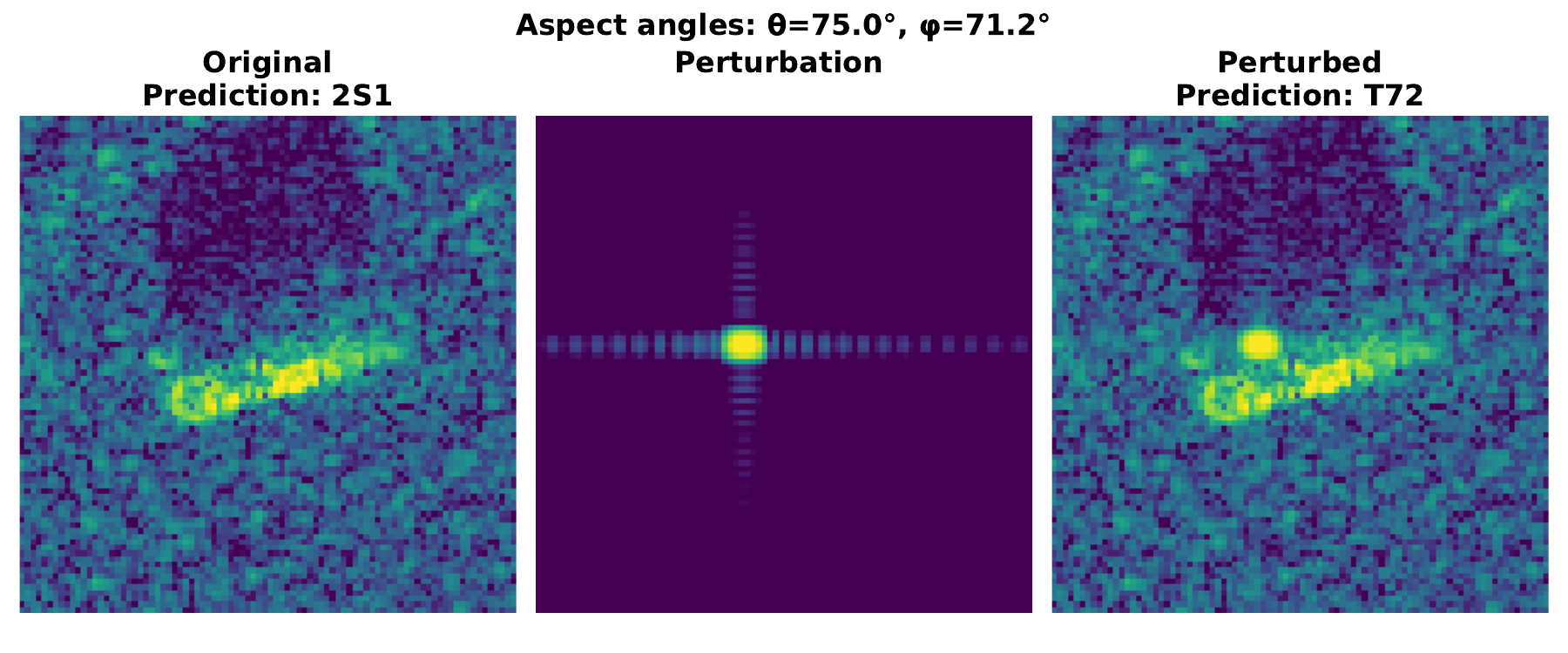}\label{fig:example_1}}
    \hfill
    \subfloat[]{\includegraphics[width=0.49\linewidth]{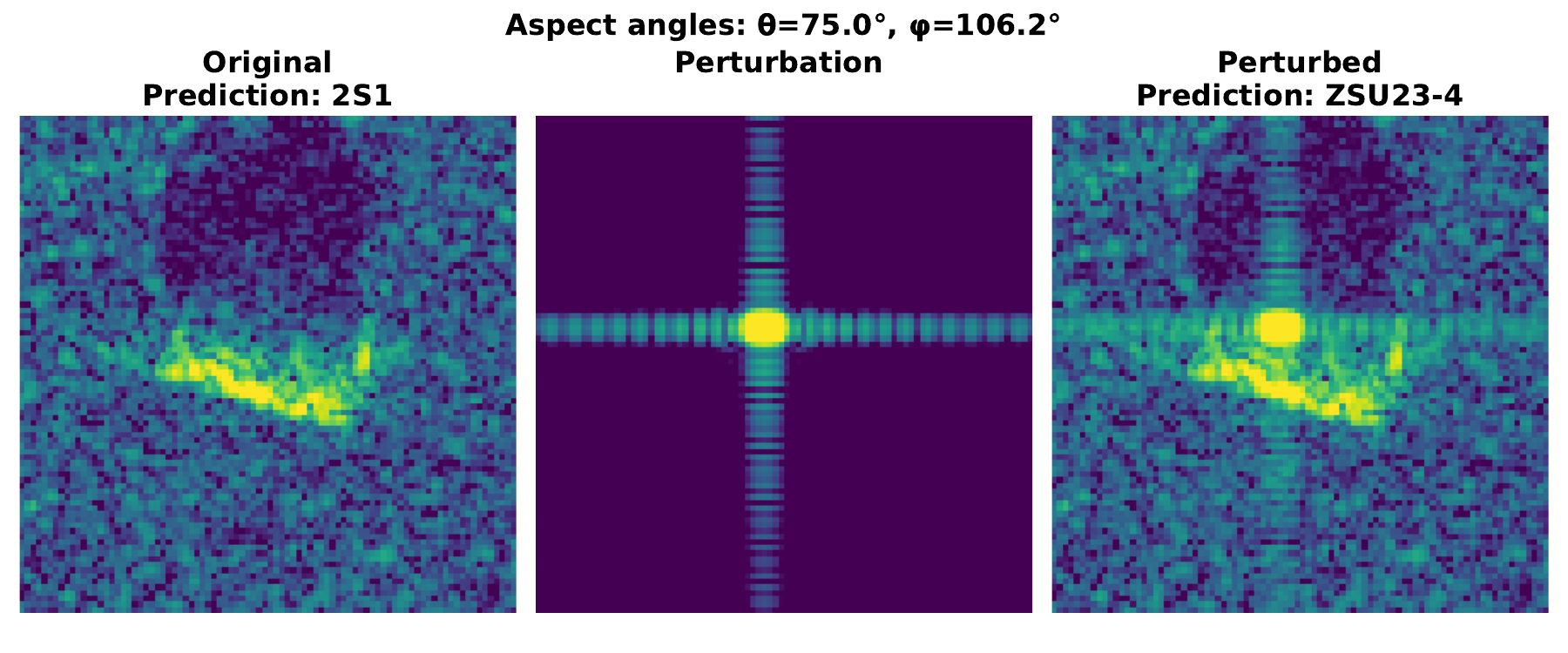}\label{fig:example_2}}
    \\
    \subfloat[]{\includegraphics[width=0.49\linewidth]{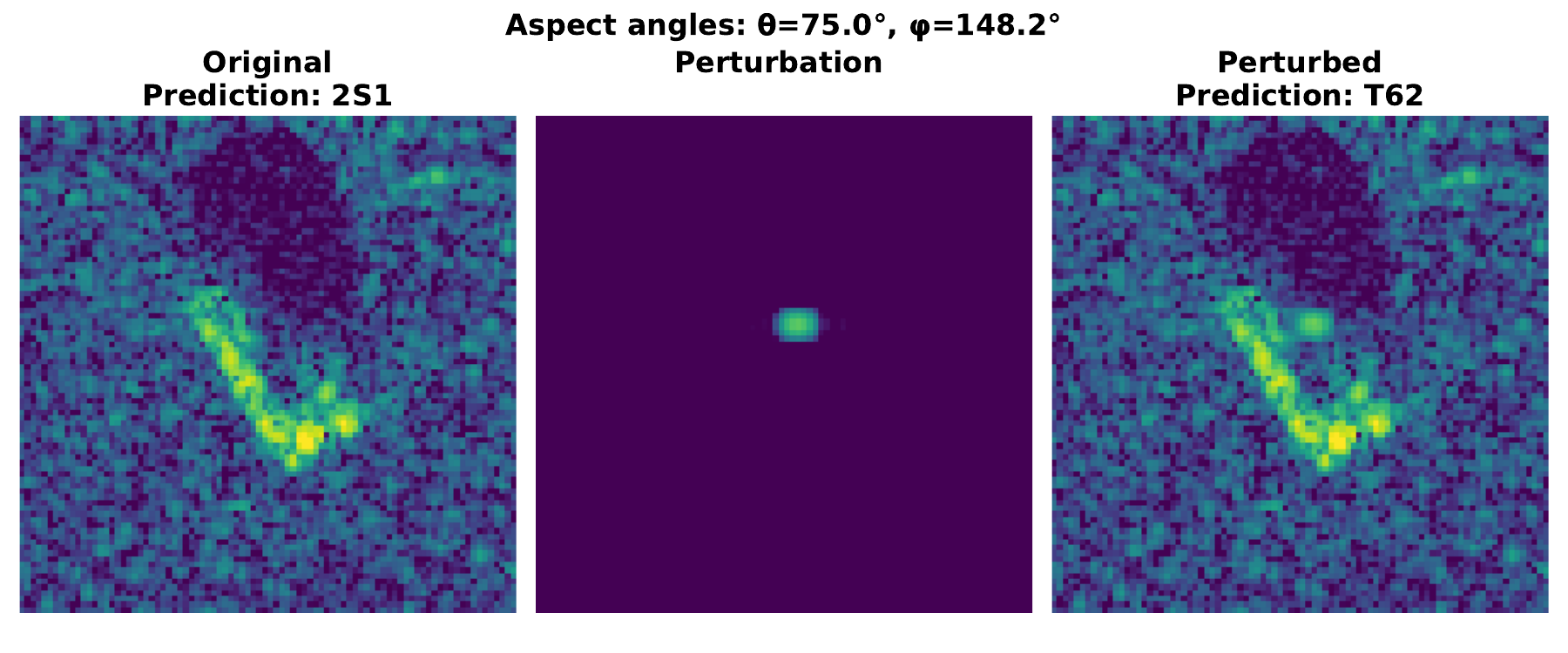}\label{fig:example_3}}
    \hfill
    \subfloat[]{\includegraphics[width=0.49\linewidth]{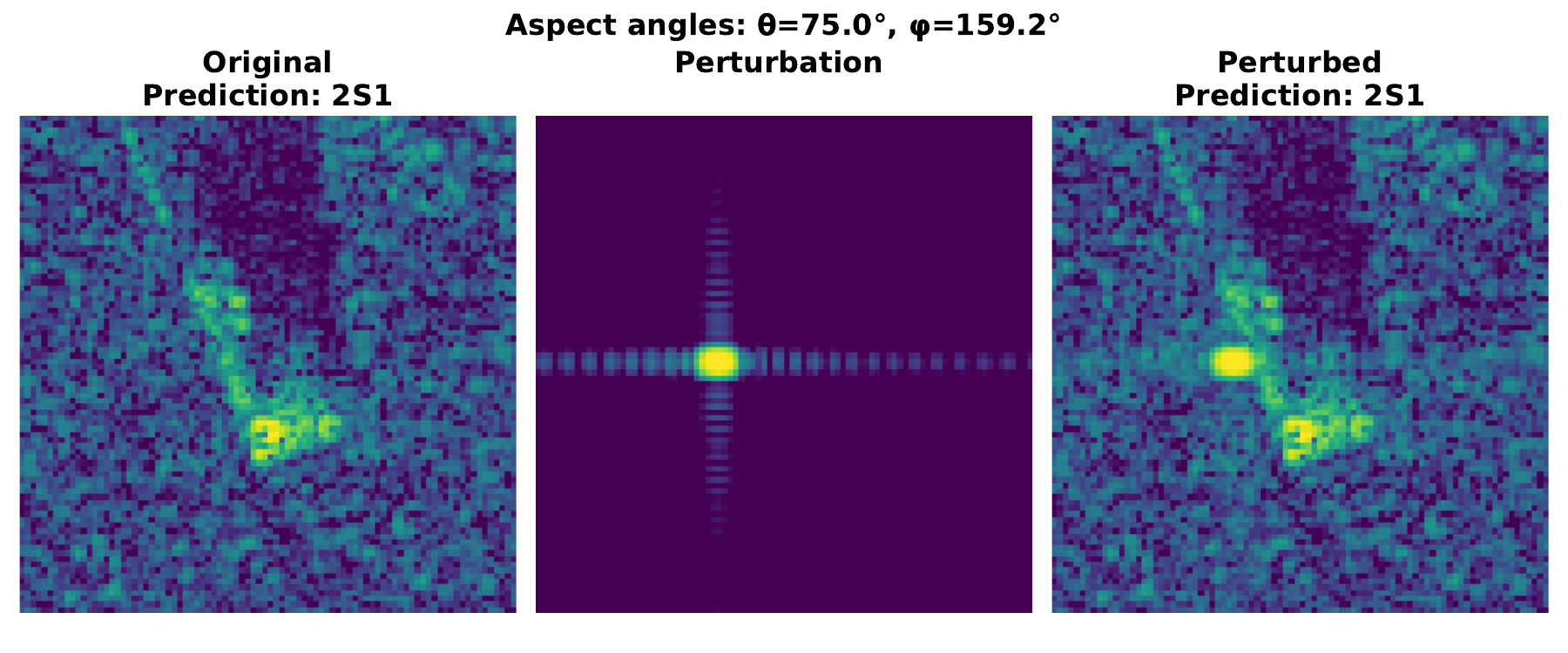}\label{fig:example_4}}
    \\
    \subfloat[]{\includegraphics[width=0.49\linewidth]{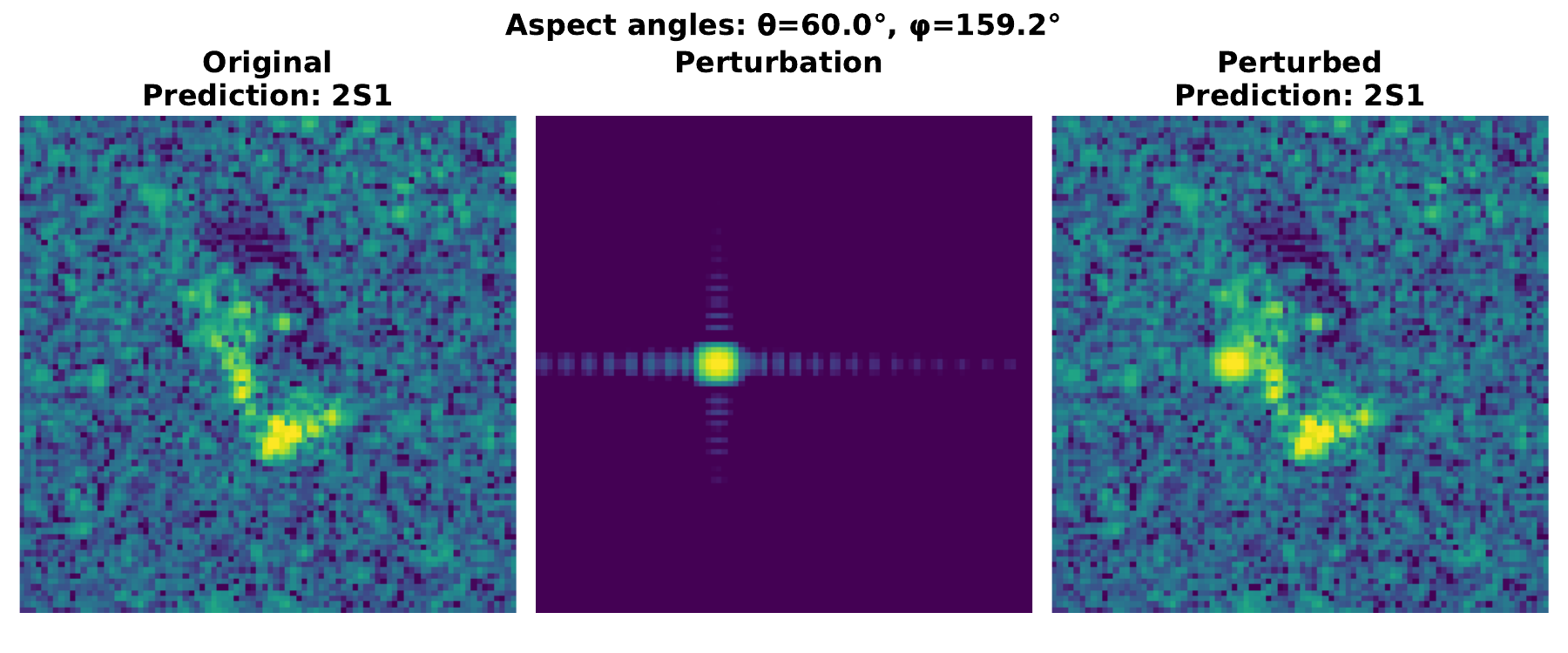}\label{fig:example_5}}
    \hfill
    \subfloat[]{\includegraphics[width=0.49\linewidth]{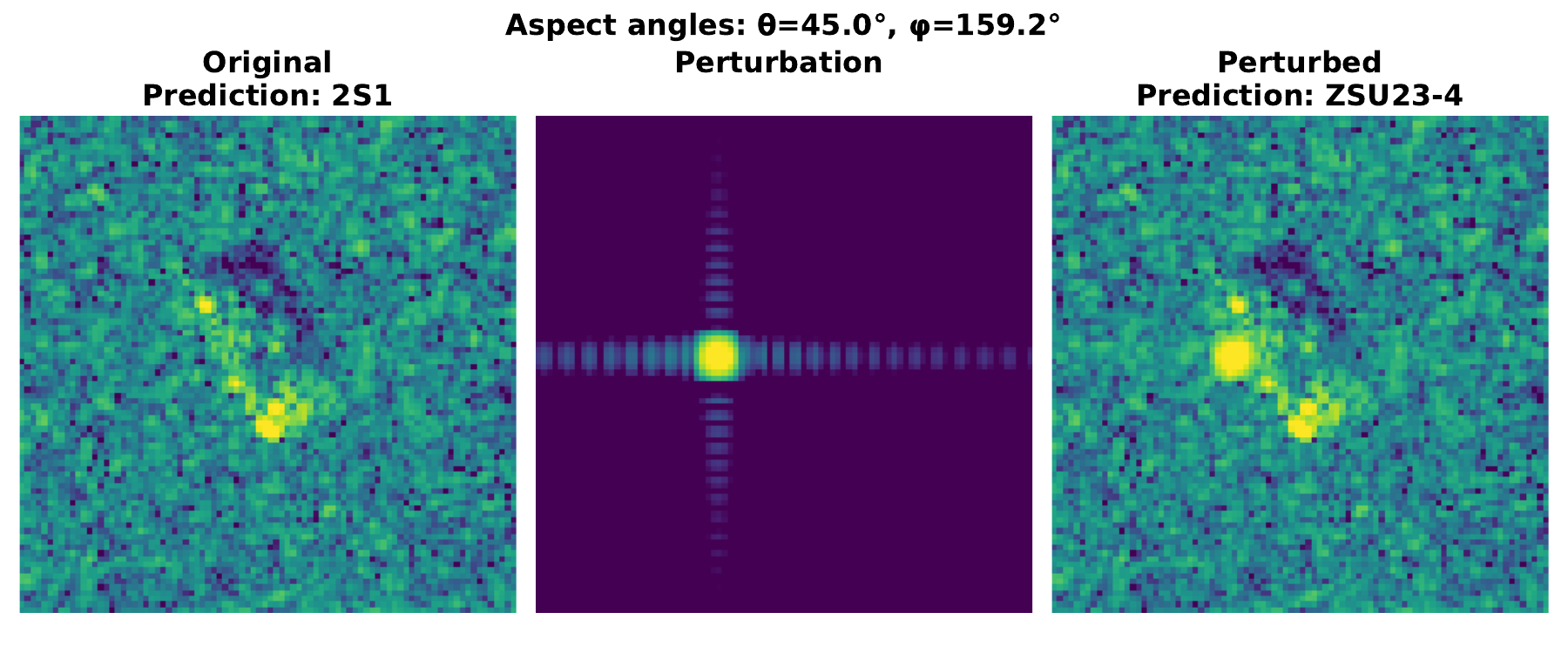}\label{fig:example_6}}
    \caption{The original scene, the perturbation parameterized as per \Cref{tab:perturbation}, and the perturbed scene observed from different aspect angles: (a) When $\theta^a = 75^\circ$, $\phi^a = 71.2^\circ$, the second reflector has its boresight at azimuth angle $\phi_2=106.3^\circ$, making it visible over the azimuth range $\phi^a \in [61.3^\circ, 151.3^\circ]$. The reflector creates a bright spot, successfully deceiving the target model, while the remaining reflectors are outside their visibility ranges and thus not visible. (b) When $\theta^a = 75^\circ$, $\phi^a = 106.2^\circ$, reflector brightness approaches its peak as the reflector is viewed near its boresight. (c) When $\theta^a = 75^\circ$, $\phi^a = 148.2^\circ$, reflector brightness diminishes as the reflector is observed near the edge of its visibility range. (d) When $\theta^a = 75^\circ$, $\phi^a = 159.2^\circ$ ($11^\circ$ apart from before in azimuth), since each reflector is oriented toward a different quadrant, visibility transitions smoothly from one reflector to the next. Here, the second reflector has dropped out of view while the third has become visible. (e) When $\theta^a = 60^\circ$, $\phi^a = 159.2^\circ$, reflector brightness shows little variation from before. (f) When $\theta^a = 45^\circ$, $\phi^a = 159.2^\circ$, reflector brightness again shows little variation from before.}
    \label{fig:examples}
\end{figure*}

\Cref{tab:perturbation} shows the result of optimizing $\Theta$ (defined in \Cref{eq:physical_parameters}) for the scene visualized in \Cref{fig:examples} specific for the 2S1 class, when four corner reflectors were used. An average fooling rate of 72.9\% was achieved on the training set and 71.0\% was achieved on the test set.

\begin{table}[ht]
    \caption{Physical properties (position and orientation of the boresight, expressed in incidence angle $\theta$ and azimuth angle $\phi$) of the adversarial reflectors for a single perturbation.}
    \label{tab:perturbation}
    \centering
    \begin{tabularx}{\linewidth}{
        >{\centering\arraybackslash}X
        >{\centering\arraybackslash}X 
        >{\centering\arraybackslash}X
        >{\centering\arraybackslash}X
        >{\centering\arraybackslash}X
    }
    \toprule
        Reflector & $x$-position (m) & $y$-position (m) & $\theta$ & $\phi$\\ \midrule
        1 & $ 0.31$ & $-2.91$ & $66.3^\circ$ & $ 16.3^\circ$ \\
        2 & $-1.62$ & $-1.80$ & $65.0^\circ$ & $106.3^\circ$ \\
        3 & $-1.55$ & $ 3.18$ & $69.2^\circ$ & $196.3^\circ$ \\
        4 & $-0.73$ & $-2.50$ & $75.0^\circ$ & $286.3^\circ$ \\
        \bottomrule
    \end{tabularx}
\end{table}

\Cref{fig:example_1} to \Cref{fig:example_3} visualize how one reflector's brightness waxes and wanes with the azimuth aspect angle. \Cref{fig:example_3} to \Cref{fig:example_4} visualize how as one reflector goes out of range, the other reflector takes its place. \Cref{fig:example_4} to \Cref{fig:example_6} visualize how for a fixed azimuth aspect angle, the \ac{SAR}-facing reflector remains visible over a range of incidence aspect angles. A supplementary video showing the full viewing sequence is provided online \cite{saaipaa-demo}. Taken together, these examples demonstrate how a single perturbation achieves continuous visibility across the full viewing domain, while also highlighting the variations in image appearance as a function of the aspect angles.

\subsubsection{Ablation with four-quadrant randomized perturbations}

To determine whether the attack performance stems from a four-quadrant coverage by the reflectors, all physical parameters $\Theta$ of the reflectors were randomly sampled, while keeping adjacent reflectors to be 90 degrees apart in the azimuth to ensure full azimuth coverage. Under these conditions, the average fooling rate dropped to $7.4\%$, compared to the previously achieved $69.1\%$. This substantial performance gap confirms that the efficacy of the attack does not stem merely from introducing conspicuous scattering centers, but from exploiting inherent adversarial vulnerabilities in the target model. Consequently, the proposed optimization framework provides a decisive advantage over na\"ive reflector configurations.

\subsubsection{Impact of unconstrained azimuth optimization}

To validate the necessity of the rigid constraint of forcing the reflectors azimuth boresight to be equally spaced \\ ($\phi_i = \left( \phi_1 + 2\pi(i-1)/m\right)$), we evaluate an ablated variant where each reflector's azimuth angle $\phi_m \in [0, 2\pi]$ is optimized independently. Removing the relative spacing constraint degrades the average fooling rate from $69.1\%$ to $47.3\%$. Expanding the azimuth search space increases optimization complexity, thereby increasing susceptibility to getting trapped in local optima, while failing to guarantee full azimuth coverage

\subsubsection{Training under limited training data}\label{sec:generalizability}

\begin{figure}[ht]
    \centering
    \includegraphics[width=0.95\linewidth]{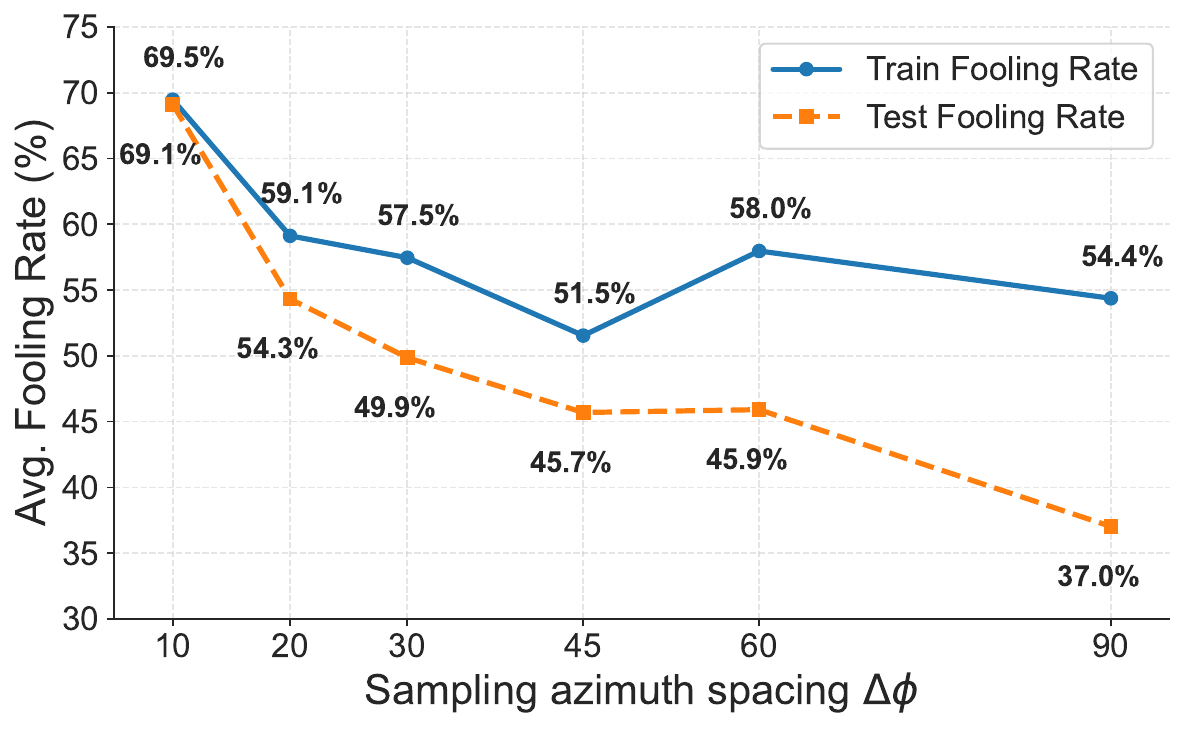}
    \caption{Average fooling rates achieved on the train and test set using various training azimuth spacing.}
    \label{fig:experiment_4}
\end{figure}

Aspect angle invariance is enabled by training physical perturbations across a representative set of observation angles. However, expanding the training set incurs a high computational cost in image synthesis and optimization. Therefore, we investigate this performance–overhead trade-off by evaluating attacks optimized across datasets sampled at varying azimuth angle spacings, the resulting average fooling rates are shown in \Cref{fig:experiment_4}. As the training azimuth spacing increases, both training and test fooling rates decrease, and the train–test gap widens. Nevertheless, even with very coarse sampling (azimuth spacing of $90^\circ$), a nontrivial fooling rate of $37.0\%$ was obtained. 


\subsubsection{Impact of Reflector Count}

We evaluate the trade-off between attack efficacy and physical complexity by varying the total number of deployed corner reflectors $m$. Under a highly constrained scenario with $m = 3$ reflectors, and thus spaced $120^\circ$ apart in azimuth, \ac{SAIPAS} achieves a fooling rate of $40.7\%$, demonstrating efficacy even with a limited angular coverage. Scaling the physical budget to $m = 8$ reflectors, ensuring at least two reflectors remain visible from any azimuth aspect angle, increases the average fooling rate to $82.2\%$, compared to $69.1\%$ for $m = 4$. These empirical results demonstrate a clear operational trade-off: increasing $m$ enhances efficacy, but demands greater physical deployment effort, and reduced physical stealth.

\subsubsection{Transferability to other models}

\begin{figure}[ht]
    \centering
    \includegraphics[width=0.95\linewidth]{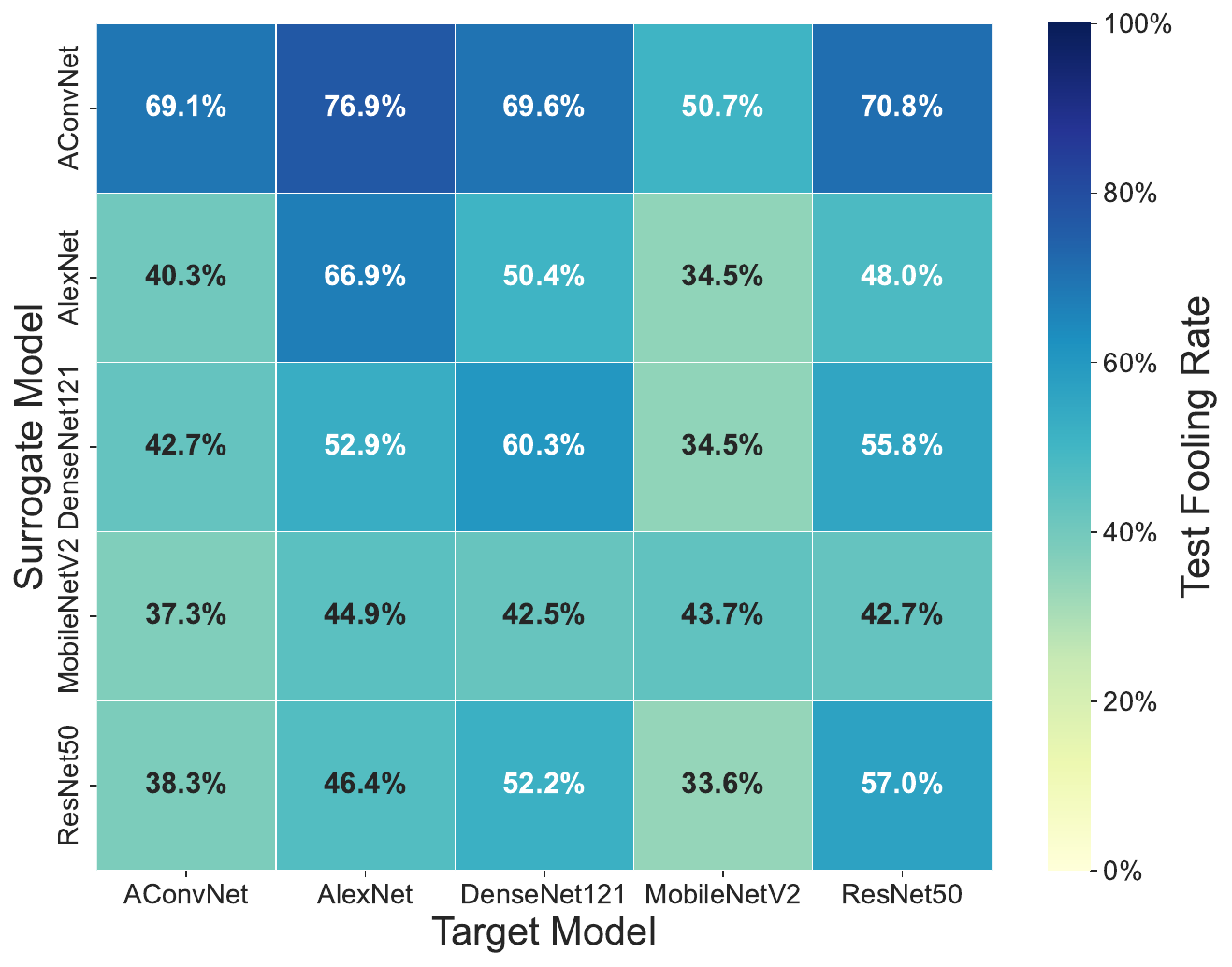}
    \caption{Average fooling rate when perturbation trained on a surrogate model was tested on a target model.}
    \label{fig:experiment_6}
\end{figure}

Even without full knowledge of the target model, adversarial attacks can succeed, by exploiting the transferability of \acp{AE}. In  this black-box scenario, the attacker trains perturbations on a surrogate model and applies them to an unknown target model, such that highly transferable perturbations induce misclassifications. To evaluate transferability, the average fooling rates for all surrogate–target model pairs were measured, as summarized in \Cref{fig:experiment_6}, where diagonal entries correspond to white-box attacks and off-diagonal entries correspond to black-box attacks.

While AConvNet surrogates demonstrate high transferability, surpassing white-box performance when targeting AlexNet ($76.9\%$), other surrogate architectures exhibit performance drops in black-box settings. MobileNetV2 consistently proves the most resistant target across all surrogates. These results demonstrate the feasibility of black-box attacks using compact surrogates, while underscoring the need for dedicated transferability-enhancing techniques. Methods for improving transferability will be investigated in future work. 

\subsubsection{Partial knowledge of the aspect angles}

So far, the attacker was assumed to have no knowledge of the aspect angles. In scenarios where partial trajectory information is available, the attacker can leverage bounded aspect angle estimates $(\phi^{\text{est}}, \theta^{\text{est}})$ satisfying:
\begin{equation}
    |\phi^{\text{est}} - \phi^a| \leq \Delta, \quad |\theta^{\text{est}} - \theta^a| \leq \Delta,
\end{equation}
where $\Delta$ denotes the aspect angle uncertainty bound. For $\Delta \leq 45^\circ$, a single corner reflector ($m=1$) suffices to cover the full potential viewing sector. In this setup, the reflector's boresight is oriented directly toward the estimated angles ($\phi_1 = \phi^{\text{est}}, \theta_1 = \theta^{\text{est}}$), leaving only its spatial position $(x_1, y_1)$ to be optimized. The training set $\mathcal{G}$ comprises a single image corresponding to $(\phi^{\text{est}}, \theta^{\text{est}})$, while the test set encompasses all target samples within the uncertainty bound $\Delta$.

\begin{figure}[ht]
    \centering
    \includegraphics[width=0.95\linewidth]{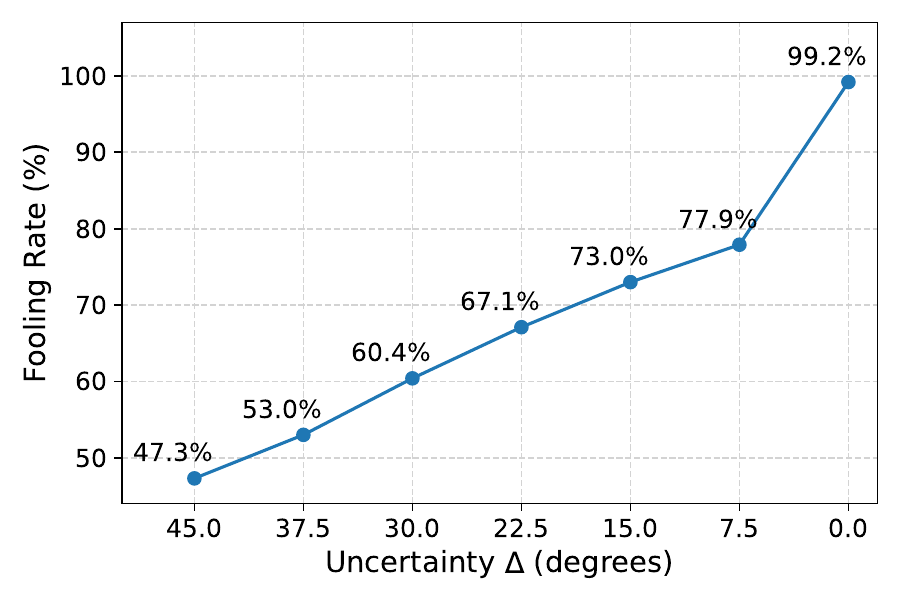}
    \caption{Average fooling rate per uncertainty $\Delta$.}
    \label{fig:experiment_8_fooling_rates}
\end{figure}

The \acp{AE} were trained using a population size of $40$ over $15$ iterations. The resulting average fooling rates are summarized in \Cref{fig:experiment_8_fooling_rates}. The average fooling rates were found to increase as the uncertainty decreased. Even under the largest tested uncertainty of $\Delta = 45^\circ$, corresponding to the full viewing range of the corner reflector, a nontrivial average fooling rate of 47.3\% was achieved.

For the case $\Delta = 0$, where the attacker had full knowledge of the aspect angles, the \ac{AE} was evaluated solely on the training sample, yielding rates as high as 99.2\%. This $\Delta = 0$ setting is the only scenario in which the assumed attacker model coincides with that adopted in prior studies \cite{peng2022scattering, xie2024migaa, zhang2024physically, ma2025sar-paa}, which implicitly restrict optimization and evaluation to a single aspect angle. 

Direct numerical comparison of raw fooling rates between \ac{SAIPAS} and existing physical \ac{SAR} adversarial attacks (e.g., \cite{peng2022scattering, xie2024migaa, zhang2024physically, ma2025sar-paa}) are omitted, since it presents a fundamental methodological mismatch. In standard digital adversarial machine learning, baseline algorithms share identical perturbation constraints, allowing direct metrics like fooling rate to serve as valid performance indicators. In physical adversarial attacks, however, frameworks vary drastically in how physical perturbations are modeled, constrained, and rendered into the \ac{SAR} image domain. Trivially high fooling rates may be achieved at the expense of realism.

\subsection{Limitations of the Simulations}

While the underlying forward electromagnetic scattering model has demonstrated strong agreement with experimental data~\cite{polycarpou1995radar}, several key limitations of our simulation framework should be noted. First, the assumed linear superposition of the scattering centers produced by the corner reflectors with the \ac{SAR} returns neglects higher-order multiple scattering interactions between the target hull and the reflectors. Second, radar shadowing and geometric occlusions were not modeled, which may overestimate the aspect angle coverage of the reflectors in occluded sectors. Finally, the attack remains validated strictly within the simulated domain; physical execution introduces unmodeled non-idealities, such as deviations from the intended flight path, surface roughness, and background clutter.

\subsection{Practical Implications}

While \ac{SAIPAS} advances physical \ac{SAR} adversarial attacks by eliminating the restrictive requirement of known platform aspect angles, several real-world constraints remain. First, the optimization relies on accurate estimates of the \ac{SAR} platform's system specifications. Furthermore, the model assumes precise spatial positioning and orientation for each reflector. Additional limitations stem from target geometry and mobility. The optimization space for the position does not enforce target-specific physical boundaries, the computed locations may prove operationally prohibitive by obstructing dynamic components, optical sensors, or structural load points on the host vehicle. Finally, the optimized reflector configuration is defined relative to the target object. Thus, if the target moves or shifts relative to its background, the perturbation loses optimality, necessitating manual remounting and realignment.
    
 \section{Conclusion}

This work is motivated by the confluence of three technological developments: (1) the proliferation of space-based \ac{SAR} systems due to their all-time, all-weather remote sensing capabilities; (2) the maturing application of \ac{ML} to \ac{SAR}-based \ac{ATR}; and (3) the deluge of discoveries in adversarial \ac{ML} threatening \ac{ML} applications, including \ac{SAR}-based \ac{ATR} systems. The increasing importance of \ac{SAR} \ac{ATR}, combined with the increasing potency of attacks against \ac{ML} applications, motivates research into novel attack mechanisms, and correspondingly defence mechanisms.

In this paper, we presented the \acf{SAIPAS}, a physics-driven framework for evaluating simulated physical adversarial threats against \ac{SAR} \ac{ATR} models. Using the framework, an attacker equipped with a fixed number of reflectors of arbitrary size can determine an adversarially effective configuration to fool a target model observing the scene from unknown aspect angles. Dispensing with the restrictive assumption of known aspect knowledge, our approach relaxes a fundamental constraint of existing literature to establish a significantly less restrictive and more plausible threat model.

The rigorous physical modeling offers explicit physical interpretability and increased modeling fidelity compared to prior frameworks, while maintaining high adversarial efficacy. When a single corner reflector is visible at any azimuth aspect angle, the attack achieves an average fooling rate of 69.1\%, compared to 7.4\% for randomized configurations with identical angular coverage. The attack remains effective even when the physical perturbation is optimized using a limited number of training samples, achieving a fooling rate of 37.0\% with training samples 90$^\circ$ apart in azimuth. In the scenario where the aspect angles are fully known, the fooling rate rises to 99.2\%. 

Future work includes improving the transferability of \ac{SAIPAS}, to increase the effectiveness of attacking unknown (black-box) target models. 

\section*{Declaration of Competing Interest}

The authors declare that they have no known competing financial interests or personal relationships that could have appeared to influence the work reported in this paper.

\section*{Acknowledgment}

This material is based upon work supported by the Air Force Office of Scientific Research under award number FA2386-23-1-4082. Isar Lemeire is also supported by the Australian Government through the Research Training Program international (RTPi) Scholarships program. The \ac{MSTAR} dataset was made available by \ac{DARPA} and \ac{AFRL}.




\bibliographystyle{elsarticle-harv}
\bibliography{IEEEabrv,sar-atr,cv-aml,sar-aml,transfer,specifications, models}

@STRING{IEEE_J_AES        = "{IEEE} Trans. Aerosp. Electron. Syst."}

@STRING{IEEE_J_NNLS       = "{IEEE} Trans. Neural Netw. Learn. Syst."}

@STRING{IEEE_J_IP         = "{IEEE} Trans. Image Process."}

@STRING{IEEE_J_PAMI       = "{IEEE} Trans. Pattern Anal. Mach. Intell."}

@STRING{IEEE_J_GRSL       = "{IEEE} Geosci. Remote Sens. Lett."}

@STRING{IEEE_J_GRS        = "{IEEE} Trans. Geosci. Remote Sens."}

@STRING{IEEE_J_STARS      = "{IEEE} J. Sel. Topics Appl. Earth Observ. Remote Sens."}

@STRING{IEEE_J_AWPL       = "{IEEE} Antennas Wireless Propag. Lett."}

@STRING{IEEE_J_AP         = "{IEEE} Trans. Antennas Propag."}

@INPROCEEDINGS{kennedy1995particle,
  author={Kennedy, J. and Eberhart, R.},
  booktitle={Proceedings of ICNN'95 - International Conference on Neural Networks}, 
  title={Particle swarm optimization}, 
  year={1995},
  volume={4},
  number={},
  pages={1942-1948},
  doi={10.1109/ICNN.1995.488968}
}

@article{storn1997differential,
	title = {Differential Evolution – A Simple and Efficient Heuristic for global Optimization over Continuous Spaces},
	volume = {11},
	issn = {1573-2916},
	doi = {10.1023/A:1008202821328},
	number = {4},
	journal = {Journal of Global Optimization},
	author = {Storn, Rainer and Price, Kenneth},
	month = dec,
	year = {1997},
	pages = {341--359},
}

@article{deb1995simulated,
  title={Simulated binary crossover for continuous search space},
  author={Deb, Kalyanmoy and Agrawal, Ram Bhushan},
  journal={Complex Systems},
  volume={9},
  number={2},
  pages={115--148},
  year={1995}
}

@ARTICLE{wang2004image,
  author={Zhou Wang and Bovik, A.C. and Sheikh, H.R. and Simoncelli, E.P.},
  journal=IEEE_J_IP, 
  title={Image quality assessment: from error visibility to structural similarity}, 
  year={2004},
  volume={13},
  number={4},
  pages={600-612},
  doi={10.1109/TIP.2003.819861}
}

@inproceedings{snoek2012practical,
 author = {Snoek, Jasper and Larochelle, Hugo and Adams, Ryan P},
 booktitle = {Advances in Neural Information Processing Systems},
 editor = {F. Pereira and C.J. Burges and L. Bottou and K.Q. Weinberger},
 pages = {},
 publisher = {Curran Associates, Inc.},
 title = {Practical Bayesian Optimization of Machine Learning Algorithms},
 url = {https://proceedings.neurips.cc/paper_files/paper/2012/file/05311655a15b75fab86956663e1819cd-Paper.pdf},
 volume = {25},
 year = {2012}
}

@misc{szegedy2014intriguing,
  title={Intriguing properties of neural networks}, 
  author={Christian Szegedy and Wojciech Zaremba and Ilya Sutskever and Joan Bruna and Dumitru Erhan and Ian Goodfellow and Rob Fergus},
  year={2014},
  howpublished={arXiv preprint arXiv:1312.6199},
  doi={10.48550/arXiv.1312.6199}, 
}

@inproceedings{goodfellow2015explaining,
  author={Ian J. Goodfellow and Jonathon Shlens and Christian Szegedy},
  title={Explaining and Harnessing Adversarial Examples}, 
  booktitle=ICLR,
  year={2015},
  doi={10.48550/arXiv.1412.6572}
}

@INPROCEEDINGS {moosavi-dezfooli2017universal,
author = { Moosavi-Dezfooli, Seyed-Mohsen and Fawzi, Alhussein and Fawzi, Omar and Frossard, Pascal },
booktitle = {2017 IEEE Conference on Computer Vision and Pattern Recognition (CVPR)},
title = {{ Universal Adversarial Perturbations }},
year = {2017},
volume = {},
ISSN = {1063-6919},
pages = {86-94},
doi = {10.1109/CVPR.2017.17},
publisher = {IEEE Computer Society},
address = {Los Alamitos, CA, USA},
month = jul
}

@InProceedings{selvaraju2017grad-cam,
  author = {Selvaraju, Ramprasaath R. and Cogswell, Michael and Das, Abhishek and Vedantam, Ramakrishna and Parikh, Devi and Batra, Dhruv},
  title = {{Grad-CAM}: Visual Explanations From Deep Networks via Gradient-Based Localization},
  booktitle = {Proceedings of the IEEE International Conference on Computer Vision (ICCV)},
  month = oct,
  year = {2017}
}

@misc{gartner-gen-ai,
    author = {Gartner},
    title = {Gartner Experts Answer the Top Generative {AI} Questions for Your Enterprise},
    howpublished = {Gartner Insights},
    year = 2024,
    url = {https://www.gartner.com/en/topics/generative-ai},
    note = {accessed 22 Jan 2024}
}

@article{nguyen2024physical,
	title        = {Physical Adversarial Attacks for Surveillance: A Survey},
	author       = {Nguyen, Kien and Fernando, Tharindu and Fookes, Clinton and Sridharan, Sridha},
	year         = 2024,
	journal      = IEEE_J_NNLS,
	volume       = 35,
	number       = 12,
	pages        = {17036--17056},
	doi          = {10.1109/TNNLS.2023.3321432}
}

@misc{nist.ai.100-2e2023,
  author = {Apostol Vassilev and Alina Oprea and Alie Fordyce and Hyrum Anderson},
  title = {Adversarial Machine Learning: A Taxonomy and Terminology of Attacks and Mitigations},
  howpublished = {NIST Trustworthy and Responsible AI, NIST AI 100-2e2023},
  year = 2024,
  month = jan,
  doi = {10.6028/NIST.AI.100-2e2023}
}

@article{wei2024physical,
	title        = {Physical Adversarial Attack Meets Computer Vision: A Decade Survey},
	author       = {Wei, Hui and Tang, Hao and Jia, Xuemei and Wang, Zhixiang and Yu, Hanxun and Li, Zhubo and Satoh, Shin’ichi and Van Gool, Luc and Wang, Zheng},
	year         = 2024,
	journal      = IEEE_J_PAMI,
	volume       = 46,
	number       = 12,
	pages        = {9797--9817},
	doi          = {10.1109/TPAMI.2024.3430860}
}

@inproceedings{krizhevsky2012imagenet,
     author = {Krizhevsky, Alex and Sutskever, Ilya and Hinton, Geoffrey E},
     booktitle = {Advances in Neural Information Processing Systems},
     editor = {F. Pereira and C.J. Burges and L. Bottou and K.Q. Weinberger},
     pages = {},
     publisher = {Curran Associates, Inc.},
     title = {{ImageNet Classification with Deep Convolutional Neural Networks}},
     url = {https://proceedings.neurips.cc/paper_files/paper/2012/file/c399862d3b9d6b76c8436e924a68c45b-Paper.pdf},
     volume = {25},
     year = {2012}
}

@InProceedings{he2016deep,
author = {He, Kaiming and Zhang, Xiangyu and Ren, Shaoqing and Sun, Jian},
title = {Deep Residual Learning for Image Recognition},
booktitle = CVPR,
pages = {770--778},
month = jun,
year = {2016},
doi = {10.1109/CVPR.2016.90}
}

@InProceedings{Huang2017densely,
    author = {Huang, Gao and Liu, Zhuang and van der Maaten, Laurens and Weinberger, Kilian Q.},
    title = {Densely Connected Convolutional Networks},
    booktitle = {Proceedings of the IEEE Conference on Computer Vision and Pattern Recognition (CVPR)},
    month = {July},
    year = {2017}
}

@INPROCEEDINGS{sandler2018mobilenetv2,
  author={Sandler, Mark and Howard, Andrew and Zhu, Menglong and Zhmoginov, Andrey and Chen, Liang-Chieh},
  booktitle={2018 IEEE/CVF Conference on Computer Vision and Pattern Recognition}, 
  title={{MobileNetV2}: Inverted Residuals and Linear Bottlenecks}, 
  year={2018},
  volume={},
  number={},
  pages={4510-4520},
  doi={10.1109/CVPR.2018.00474}
}

@ARTICLE{chen2021empirical,
  author={Chen, Li and Xu, Zewei and Li, Qi and Peng, Jian and Wang, Shaowen and Li, Haifeng},
  journal=IEEE_J_GRS, 
  title={An Empirical Study of Adversarial Examples on Remote Sensing Image Scene Classification}, 
  year={2021},
  volume={59},
  number={9},
  pages={7419-7433},
  doi={10.1109/TGRS.2021.3051641}
}

@Article{du2021adversarial,
  AUTHOR = {Du, Chuan and Zhang, Lei},
  TITLE = {Adversarial Attack for {SAR} Target Recognition Based on {UNet}-Generative Adversarial Network},
  JOURNAL = {Remote Sensing},
  VOLUME = {13},
  YEAR = {2021},
  NUMBER = {21},
  ARTICLE-NUMBER = {4358},
  ISSN = {2072-4292},
  DOI = {10.3390/rs13214358}
}

@ARTICLE{li2021adversarial,
  author={Li, Haifeng and Huang, Haikuo and Chen, Li and Peng, Jian and Huang, Haozhe and Cui, Zhenqi and Mei, Xiaoming and Wu, Guohua},
  journal=IEEE_J_STARS, 
  title={Adversarial Examples for {CNN}-Based {SAR} Image Classification: An Experience Study}, 
  year={2021},
  volume={14},
  pages={1333-1347},
  doi={10.1109/JSTARS.2020.3038683}
}

@ARTICLE{du2022fast,
  author={Du, Chuan and Huo, Chaoying and Zhang, Lei and Chen, Bo and Yuan, Yijun},
  journal=IEEE_J_GRSL, 
  title={Fast {C\&W}: A Fast Adversarial Attack Algorithm to Fool {SAR} Target Recognition With Deep Convolutional Neural Networks}, 
  year={2022},
  volume={19},
  number={},
  pages={1-5},
  doi={10.1109/LGRS.2021.3058011}
}

@ARTICLE{peng2022scattering,
  author={Peng, Bowen and Peng, Bo and Zhou, Jie and Xie, Jianyue and Liu, Li},
  journal=IEEE_J_GRS, 
  title={Scattering Model Guided Adversarial Examples for {SAR} Target Recognition: Attack and Defense}, 
  year={2022},
  volume={60},
  number={},
  pages={1-17},
  doi={10.1109/TGRS.2022.3213305}
}

@ARTICLE{peng2022speckle,
  author={Peng, Bowen and Peng, Bo and Zhou, Jie and Xia, Jingyuan and Liu, Li},
  journal=IEEE_J_GRSL, 
  title={Speckle-Variant Attack: Toward Transferable Adversarial Attack to {SAR} Target Recognition}, 
  year={2022},
  volume={19},
  number={},
  pages={1-5},
  doi={10.1109/LGRS.2022.3184311}
}

@INPROCEEDINGS{chen2023positive,
  author={Chen, Yuzhou and Du, Jiawei and Yang, Yang and Sun, Changfeng},
  booktitle={2023 IEEE 3rd International Conference on Electronic Technology, Communication and Information (ICETCI)}, 
  title={Positive Weighted Feature Attack: Toward Transferable Adversarial Attack to {SAR} Target Recognition}, 
  year={2023},
  pages={93-98},
  doi={10.1109/ICETCI57876.2023.10176719}
}

@Article{du2023tan,
AUTHOR = {Du, Meng and Sun, Yuxin and Sun, Bing and Wu, Zilong and Luo, Lan and Bi, Daping and Du, Mingyang},
TITLE = {{TAN: A Transferable Adversarial Network for DNN-Based UAV SAR Automatic Target Recognition Models}},
JOURNAL = {Drones},
VOLUME = {7},
YEAR = {2023},
NUMBER = {3},
ARTICLE-NUMBER = {205},
ISSN = {2504-446X},
DOI = {10.3390/drones7030205}
}

@InProceedings{peng2023low,
  author="Peng, Bo and Peng, Bowen and Zhou, Jie and Huang, Xichen and Meng, Lingxin and Gao, Xunzhang",
  title="Low-Frequency Features Optimization for Transferability Enhancement in Radar Target Adversarial Attack",
  booktitle="Artificial Neural Networks and Machine Learning -- ICANN 2023",
  year="2023",
  publisher="Springer Nature Switzerland",
  address="Cham",
  pages="115--129",
  isbn="978-3-031-44192-9",
  doi="10.1007/978-3-031-44192-9_10"
}

@ARTICLE{qin2023scma,
  author={Qin, Weibo and Long, Bo and Wang, Feng},
  journal=IEEE_J_GRSL, 
  title={{SCMA}: A Scattering Center Model Attack on {CNN-SAR} Target Recognition}, 
  year={2023},
  volume={20},
  pages={1-5},
  doi={10.1109/LGRS.2023.3253189}
}

@ARTICLE{xia2023sar-pega,
  author={Xia, Weijie and Liu, Zhe and Li, Yi},
  journal=IEEE_J_AES, 
  title={{SAR-PeGA: A Generation Method of Adversarial Examples for SAR Image Target Recognition Network}}, 
  year={2023},
  volume={59},
  number={2},
  pages={1910-1920},
  doi={10.1109/TAES.2022.3206261}
}

@INPROCEEDINGS{yu2023sar,
  author={Yu, Yameng and Zou, Haiyan and Zhang, Fan},
  booktitle={IGARSS 2023 - 2023 IEEE International Geoscience and Remote Sensing Symposium}, 
  title={{SAR Sticker: An Adversarial Image Patch that can Deceive {SAR ATR} Deep Model}},
  year={2023},
  pages={7050-7053},
  doi={10.1109/IGARSS52108.2023.10282390}
}

@ARTICLE{zhou2023attributed,
  author={Zhou, Junfan and Feng, Sijia and Sun, Hao and Zhang, Linbin and Kuang, Gangyao},
  journal=IEEE_J_GRSL, 
  title={Attributed Scattering Center Guided Adversarial Attack for {DCNN SAR} Target Recognition}, 
  year={2023},
  volume={20},
  pages={1-5},
  doi={10.1109/LGRS.2023.3235051},
}

@Article{luo2024sar-patt,
    AUTHOR = {Luo, Binyan and Cao, Hang and Cui, Jiahao and Lv, Xun and He, Jinqiang and Li, Haifeng and Peng, Chengli},
    TITLE = {{SAR-PATT: A Physical Adversarial Attack for SAR Image Automatic Target Recognition}},
    JOURNAL = {Remote Sensing},
    VOLUME = {17},
    YEAR = {2025},
    NUMBER = {1},
    ARTICLE-NUMBER = {21},
    ISSN = {2072-4292},
    DOI = {10.3390/rs17010021}
}

@INPROCEEDINGS{xie2024migaa,
  author={Xie, Jianyue and Peng, Bo and Lu, Zhengzhi and Zhou, Jie and Peng, Bowen},
  booktitle={2024 9th International Conference on Computer and Communication Systems (ICCCS)}, 
  title={{MIGAA: A Physical Adversarial Attack Method against SAR Recognition Models}},
  year={2024},
  pages={309-314},
  doi={10.1109/ICCCS61882.2024.10602913}
}

@ARTICLE{zhang2024physically,
  author={Zhang, Fan and Yu, Yameng and Ma, Fei and Zhou, Yongsheng},
  journal={IEEE Journal of Selected Topics in Applied Earth Observations and Remote Sensing}, 
  title={A Physically Realizable Adversarial Attack Method Against {SAR} Target Recognition Model}, 
  year={2024},
  volume={17},
  number={},
  pages={11943-11957},
  doi={10.1109/JSTARS.2024.3420690}
}

@article{peng2024towards,
    title = {Towards assessing the synthetic-to-measured adversarial vulnerability of SAR ATR},
    journal = {ISPRS Journal of Photogrammetry and Remote Sensing},
    volume = {214},
    pages = {119-134},
    year = {2024},
    issn = {0924-2716},
    doi = {https://doi.org/10.1016/j.isprsjprs.2024.06.004},
    url = {https://www.sciencedirect.com/science/article/pii/S092427162400234X},
    author = {Bowen Peng and Bo Peng and Jingyuan Xia and Tianpeng Liu and Yongxiang Liu and Li Liu},
}

@ARTICLE{ma2025sar-paa,
  author={Ma, Yanjing and Pei, Jifang and Huo, Weibo and Zhang, Yin and Huang, Yulin and Chen, Keyang and Yang, Jianyu},
  journal={IEEE Transactions on Aerospace and Electronic Systems}, 
  title={{SAR-PAA: A Physically Adversarial Attack Approach Against SAR Intelligent Target Recognition}}, 
  year={2025},
  volume={61},
  number={2},
  pages={1377-1393},
  doi={10.1109/TAES.2024.3456750}
}

@misc{saaipaa-demo,
  author       = {Isar Lemeire},
  title        = {{SAIPAS demo}},
  howpublished = {\url{https://youtu.be/NhpHHQeXaEQ?si=9SRsKGutzPCrxv7J}},
  note         = {Accessed: 6 Oct 2026}
}

@article{zhang2026towards,
    title = {Towards data-free and model-agnostic black-box attacks against SAR-ATR via dual-diversity augmentation},
    journal = {ISPRS Journal of Photogrammetry and Remote Sensing},
    volume = {238},
    pages = {282-299},
    year = {2026},
    issn = {0924-2716},
    doi = {https://doi.org/10.1016/j.isprsjprs.2026.05.006},
    url = {https://www.sciencedirect.com/science/article/pii/S0924271626002479},
    author = {Xiaoxuan Zhang and Yang Li and Zhi-Liang Zhao},
}

@article{keller1962geometrical,
author = {Joseph B. Keller},
journal = {J. Opt. Soc. Am.},
number = {2},
pages = {116--130},
publisher = {Optica Publishing Group},
title = {Geometrical Theory of Diffraction},
volume = {52},
month = feb,
year = {1962},
doi = {10.1364/JOSA.52.000116},
}

@ARTICLE{raney1994precision,
  author={Raney, R.K. and Runge, H. and Bamler, R. and Cumming, I.G. and Wong, F.H.},
  journal={IEEE Transactions on Geoscience and Remote Sensing}, 
  title={Precision {SAR} processing using chirp scaling}, 
  year={1994},
  volume={32},
  number={4},
  pages={786-799},
  doi={10.1109/36.298008}
}

@article{polycarpou1995radar,
	author = {Polycarpou, Anastasis C. and Balanis, Constantine A. and Birtcher, Craig R.},
	title = {Radar cross section of trihedral corner reflectors using {PO} and {MEC}},
	journal = {Annales Des Télécommunications},
	volume = {50},
	issn = {1958-9395},
	doi = {10.1007/BF02995750},
	number = {5},
	month = may,
	year = {1995},
	pages = {510--516},
}

@ARTICLE{potter1997attributed,
  author={Potter, L.C. and Moses, R.L.},
  journal=IEEE_J_IP, 
  title={Attributed scattering centers for {SAR ATR}}, 
  year={1997},
  volume={6},
  number={1},
  pages={79-91},
  doi={10.1109/83.552098}
}

@inproceedings{ross1998standard,
author = {Timothy D. Ross and Steven W. Worrell and Vincent J. Velten and John C. Mossing and Michael Lee Bryant},
title = {{Standard SAR ATR evaluation experiments using the MSTAR public release data set}},
volume = {3370},
booktitle = {Algorithms for Synthetic Aperture Radar Imagery V},
editor = {Edmund G. Zelnio},
organization = {International Society for Optics and Photonics},
publisher = {SPIE},
pages = {566 -- 573},
year = {1998},
doi = {10.1117/12.321859},
note = {Dataset at \url{https://www.sdms.afrl.af.mil/index.php?collection=mstar}}
}

@inproceedings{akyildiz1999scattering,
author = {Yeliz Akyildiz and Randolph L. Moses},
title = {{Scattering center model for SAR imagery}},
volume = {3869},
booktitle = {SAR Image Analysis, Modeling, and Techniques II},
editor = {Francesco Posa},
organization = {International Society for Optics and Photonics},
publisher = {SPIE},
pages = {76 -- 85},
year = {1999},
doi = {10.1117/12.373151},
}

@ARTICLE{gerry1999parametric,
  author={Gerry, M.J. and Potter, L.C. and Gupta, I.J. and Van Der Merwe, A.},
  journal=IEEE_J_AP,
  title={A parametric model for synthetic aperture radar measurements},
  year={1999},
  volume={47},
  number={7},
  pages={1179-1188},
  doi={10.1109/8.785750}
}

@book{balanis2012advanced,
  title     = {Advanced Engineering Electromagnetics},
  author    = {Balanis, Constantine A.},
  year      = {2012},
  publisher = {John Wiley \& Sons},
  edition   = {2nd},
  isbn      = {978-0470589489}
}

@book{knott2012radar,
  title={Radar cross section measurements},
  author={Knott, Eugene F.},
  year={2012},
  publisher={Springer New York, NY},
  doi={10.1007/978-1-4684-9904-9}
}

@INPROCEEDINGS{auer2016raysar,
  author={Auer, Stefan and Bamler, Richard and Reinartz, Peter},
  booktitle={2016 IEEE International Geoscience and Remote Sensing Symposium (IGARSS)}, 
  title={{RaySAR - 3D SAR} simulator: Now open source}, 
  year={2016},
  pages={6730-6733},
  doi={10.1109/IGARSS.2016.7730757}
}

@ARTICLE{chen2016target,
  author={Chen, Sizhe and Wang, Haipeng and Xu, Feng and Jin, Ya-Qiu},
  journal=IEEE_J_GRS,
  title={Target Classification Using the Deep Convolutional Networks for {SAR} Images}, 
  year={2016},
  volume={54},
  number={8},
  pages={4806-4817},
  doi={10.1109/TGRS.2016.2551720}
}

@ARTICLE{ding2016convolutional,
  author={Ding, Jun and Chen, Bo and Liu, Hongwei and Huang, Mengyuan},
  journal=IEEE_J_GRSL, 
  title={Convolutional Neural Network With Data Augmentation for {SAR} Target Recognition}, 
  year={2016},
  volume={13},
  number={3},
  pages={364-368},
  doi={10.1109/LGRS.2015.2513754}
}

@misc{furukawa2017deep,
    title={Deep Learning for Target Classification from {SAR} Imagery: Data Augmentation and Translation Invariance}, 
    author={Hidetoshi Furukawa},
    year={2017},
    howpublished={arXiv preprint arXiv:1708.07920},
    eprint={1708.07920},
    archivePrefix={arXiv},
    primaryClass={cs.CV},
    doi={10.48550/arXiv.1708.07920},
}

@ARTICLE{lin2017deep,
  author={Lin, Zhao and Ji, Kefeng and Kang, Miao and Leng, Xiangguang and Zou, Huanxin},
  journal={IEEE Geoscience and Remote Sensing Letters}, 
  title={Deep Convolutional Highway Unit Network for {SAR} Target Classification With Limited Labeled Training Data}, 
  year={2017},
  volume={14},
  number={7},
  pages={1091-1095},
  doi={10.1109/LGRS.2017.2698213}
}

@article{huang2018opensarship,
  author={Huang, Lanqing and Liu, Bin and Li, Boying and Guo, Weiwei and Yu, Wenhao and Zhang, Zenghui and Yu, Wenxian},
  journal=IEEE_J_STARS,
  title={{OpenSARShip}: A Dataset Dedicated to {Sentinel-1} Ship Interpretation}, 
  year={2018},
  volume={11},
  number={1},
  pages={195-208},
  doi={10.1109/JSTARS.2017.2755672}
}

@ARTICLE{shang2018sar,
  author={Shang, Ronghua and Wang, Jiaming and Jiao, Licheng and Stolkin, Rustam and Hou, Biao and Li, Yangyang},
  journal={IEEE Journal of Selected Topics in Applied Earth Observations and Remote Sensing}, 
  title={{SAR Targets Classification Based on Deep Memory Convolution Neural Networks and Transfer Parameters}}, 
  year={2018},
  volume={11},
  number={8},
  pages={2834-2846},
  doi={10.1109/JSTARS.2018.2836909}
}

@ARTICLE{wang2018synthetic,
  author={Wang, Junjie and Feng, Dejun and Xu, Letao and Hu, Weidong},
  journal=IEEE_J_AWPL, 
  title={Synthetic Aperture Radar Image Modulation Using Phase-Switched Screen}, 
  year={2018},
  volume={17},
  number={5},
  pages={911-915},
  doi={10.1109/LAWP.2018.2823079}
}

@Article{zhao2018multi,
AUTHOR = {Zhao, Pengfei and Liu, Kai and Zou, Hao and Zhen, Xiantong},
TITLE = {{Multi-Stream Convolutional Neural Network for SAR Automatic Target Recognition}},
JOURNAL = {Remote Sensing},
VOLUME = {10},
YEAR = {2018},
NUMBER = {9},
ARTICLE-NUMBER = {1473},
ISSN = {2072-4292},
DOI = {10.3390/rs10091473}
}

@ARTICLE{zhou2018sar,
  author={Zhou, Feng and Wang, Li and Bai, Xueru and Hui, Ye},
  journal={{IEEE Transactions on Geoscience and Remote Sensing}}, 
  title={{SAR ATR of Ground Vehicles Based on LM-BN-CNN}}, 
  year={2018},
  volume={56},
  number={12},
  pages={7282-7293},
  doi={10.1109/TGRS.2018.2849967}
}

@article{xie2019novel,
author = {Xie, Yinjie and Dai, Wenxin and Hu, Zhenxin and Liu, Yijing and Li, Chuan and Pu, Xuemei},
title = {A Novel Convolutional Neural Network Architecture for {SAR} Target Recognition},
journal = {Journal of Sensors},
volume = {2019},
number = {1},
pages = {1246548},
doi = {10.1155/2019/1246548},
year = {2019}
}

@article{wang2020sar,
author = {Wang, Wei and Zhang, Chengwen and Tian, Jinge and Ou, Jianping and Li, Ji},
title = {{A SAR Image Target Recognition Approach via Novel SSF-Net Models}},
journal = {Computational Intelligence and Neuroscience},
volume = {2020},
number = {1},
pages = {8859172},
doi = {10.1155/2020/8859172},
year = {2020}
}

@Article{zhang2020ls-ssdd,
AUTHOR = {Zhang, Tianwen and Zhang, Xiaoling and Ke, Xiao and Zhan, Xu and Shi, Jun and Wei, Shunjun and Pan, Dece and Li, Jianwei and Su, Hao and Zhou, Yue and Kumar, Durga},
TITLE = {{LS-SSDD-v1.0: A Deep Learning Dataset Dedicated to Small Ship Detection from Large-Scale Sentinel-1 SAR Images}},
JOURNAL = {Remote Sensing},
VOLUME = {12},
YEAR = {2020},
NUMBER = {18},
ARTICLE-NUMBER = {2997},
ISSN = {2072-4292},
DOI = {10.3390/rs12182997}
}

@ARTICLE{dong2021global,
  author={Dong, Ganggang and Liu, Hongwei},
  journal={IEEE Transactions on Cybernetics}, 
  title={{Global Receptive-Based Neural Network for Target Recognition in SAR Images}}, 
  year={2021},
  volume={51},
  number={4},
  pages={1954-1967},
  doi={10.1109/TCYB.2019.2952400}
}

@book{jansing2021introduction,
  author = {E. David Jansing},
  title = {Introduction to Synthetic Aperture Radar: Concepts and Practice},
  publisher = {McGraw-Hill Education},
  year = 2021,
  url = {https://www.accessengineeringlibrary.com/content/book/9781260458961}
}

@Article{zhang2021sar,
AUTHOR = {Zhang, Tianwen and Zhang, Xiaoling and Li, Jianwei and Xu, Xiaowo and Wang, Baoyou and Zhan, Xu and Xu, Yanqin and Ke, Xiao and Zeng, Tianjiao and Su, Hao and Ahmad, Israr and Pan, Dece and Liu, Chang and Zhou, Yue and Shi, Jun and Wei, Shunjun},
TITLE = {{SAR Ship Detection Dataset (SSDD): Official Release and Comprehensive Data Analysis}},
JOURNAL = {Remote Sensing},
VOLUME = {13},
YEAR = {2021},
NUMBER = {18},
ARTICLE-NUMBER = {3690},
ISSN = {2072-4292},
DOI = {10.3390/rs13183690}
}

@bOOK{harrison2022introduction,
  author={Harrison, Lee Andrew (Andy)},
  title={Introduction to Synthetic Aperture Radar Using Python and MATLAB\textregistered},
  publisher={Artech House},
  year={2022},
  url={https://ieeexplore.ieee.org/document/9893146}
}

@Article{li2023comprehensive,
  AUTHOR = {Li, Jianwei and Yu, Zhentao and Yu, Lu and Cheng, Pu and Chen, Jie and Chi, Cheng},
  TITLE = {A Comprehensive Survey on {SAR ATR} in Deep-Learning Era},
  JOURNAL = {Remote Sensing},
  VOLUME = {15},
  YEAR = {2023},
  NUMBER = {5},
  ARTICLE-NUMBER = {1454},
  ISSN = {2072-4292},
  DOI = {10.3390/rs15051454}
}

@Article{lin2023sived,
    AUTHOR = {Lin, Xin and Zhang, Bo and Wu, Fan and Wang, Chao and Yang, Yali and Chen, Huiqin},
    TITLE = {{SIVED: A SAR Image Dataset for Vehicle Detection Based on Rotatable Bounding Box}},
    JOURNAL = {Remote Sensing},
    VOLUME = {15},
    YEAR = {2023},
    NUMBER = {11},
    ARTICLE-NUMBER = {2825},
    ISSN = {2072-4292},
    DOI = {10.3390/rs15112825}
}

@ARTICLE{wang2023category,
  author={Wang, Chao and Ruan, Rui and Zhao, Zhicheng and Li, Chenglong and Tang, Jin},
  journal={IEEE Transactions on Geoscience and Remote Sensing}, 
  title={Category-Oriented Localization Distillation for {SAR} Object Detection and a Unified Benchmark}, 
  year={2023},
  volume={61},
  number={},
  pages={1-14},
  doi={10.1109/TGRS.2023.3291356}
}

@Article{yang2023yang,
    AUTHOR = {Yang, Xinpeng and Zhang, Qiang and Dong, Qiulei and Han, Zhen and Luo, Xiliang and Wei, Dongdong},
    TITLE = {Ship Instance Segmentation Based on Rotated Bounding Boxes for {SAR} Images},
    JOURNAL = {Remote Sensing},
    VOLUME = {15},
    YEAR = {2023},
    NUMBER = {5},
    ARTICLE-NUMBER = {1324},
    ISSN = {2072-4292},
    DOI = {10.3390/rs15051324}
}

@misc{iceye2025iceye,
  title        = {{ICEYE launches four new satellites and introduces its new Generation 4 satellite}},
  author       = {{ICEYE}},
  howpublished = {press release},
  url = {https://www.iceye.com/newsroom/press-releases/iceye-launches-four-new-satellites-and-introduces-its-new-generation-4-satellite},
  year         = {2025},
  note         = {published: 15 Mar 2025, accessed: 24 Jun 2025}
}

@ARTICLE{lang2025recent,
  author={Lang, Ping and Fu, Xiongjun and Dong, Jian and Yang, Huizhang and Yin, Junjun and Yang, Jian and Martorella, Marco},
  journal={IEEE Journal of Selected Topics in Applied Earth Observations and Remote Sensing}, 
  title={{Recent Advances in Deep-Learning-Based SAR Image Target Detection and Recognition}}, 
  year={2025},
  volume={18},
  number={},
  pages={6884-6915},
  doi={10.1109/JSTARS.2025.3543531}
}

@misc{up402025capella,
  author = {{UP42}},
  title = {Capella Space},
  month = jun,
  year = 2025,
  url = {https://docs.up42.com/data/datasets/capella-space},
  note = {last updated: 12 Jun 2025, accessed: 24 Jun 2025}
}

@ARTICLE{shannon1949communication,
  author={Shannon, C.E.},
  journal={Proceedings of the IRE}, 
  title={Communication in the Presence of Noise}, 
  year={1949},
  volume={37},
  number={1},
  pages={10-21},
  doi={10.1109/JRPROC.1949.232969}}

@inproceedings{perna2019imaging,
    author = {S. Perna and A. Natale and C. Esposito and P. Berardino and G. Palmese and R. Lanari},
    title = {{Imaging capabilities of an airborne X-band SAR based on the FMCW technology}},
    volume = {11059},
    booktitle = {Multimodal Sensing: Technologies and Applications},
    editor = {Ettore Stella},
    organization = {International Society for Optics and Photonics},
    publisher = {SPIE},
    pages = {110590G},
    year = {2019},
    doi = {10.1117/12.2527924},
    URL = {https://doi.org/10.1117/12.2527924}
}

@inproceedings{walls2014multi,
    author = {Thomas J. Walls and Michael L. Wilson and David Madsen and Mark Jensen and Stephanie Sullivan and Michael Addario and Iain Hally},
    title = {{Multi-mission, autonomous, synthetic aperture radar}},
    volume = {9077},
    booktitle = {Radar Sensor Technology XVIII},
    editor = {Kenneth I. Ranney and Armin Doerry},
    organization = {International Society for Optics and Photonics},
    publisher = {SPIE},
    pages = {907706},
    year = {2014},
    doi = {10.1117/12.2053561},
    URL = {https://doi.org/10.1117/12.2053561}
}

@INPROCEEDINGS{horn2017fsar,
  author={Horn, Ralf and Jaeger, Marc and Keller, Martin and Limbach, Markus and Nottensteiner, Anton and Pardini, Matteo and Reigber, Andreas and Scheiber, Rolf},
  booktitle={2017 18th International Radar Symposium (IRS)}, 
  title={{F-SAR - recent upgrades and campaign activities}}, 
  year={2017},
  volume={},
  number={},
  pages={1-10},
  doi={10.23919/IRS.2017.8008092}}

@inproceedings{Giovanni2004current,
    author = {Giovanni Alberti and Luigi Citarella and Luca Ciofaniello and Roberto Fusco and Giovanni Galiero and Aurelio Minoliti and Antonio Moccia and Marco Sacchettino and Giuseppe Salzillo},
    title = {{Current status of the development of an Italian airborne SAR system (MINISAR)}},
    volume = {5236},
    booktitle = {SAR Image Analysis, Modeling, and Techniques VI},
    editor = {Francesco Posa},
    organization = {International Society for Optics and Photonics},
    publisher = {SPIE},
    pages = {53 -- 59},
    year = {2004},
    doi = {10.1117/12.512224},
    URL = {https://doi.org/10.1117/12.512224}
}

@misc{d7,
  author       = {3Dwarehouse.com},
  title        = {logging bulldozer},
  howpublished = {https://3dwarehouse.sketchup.com/model/u995475b0-8d1c-422e-93ba-de911189c6bd/logging-bulldozer?hl=en},
  note         = {Accessed: 04 Nov 2024}
}

@InProceedings{zhou2018transferable,
  author="Zhou, Wen and Hou, Xin and Chen, Yongjun and Tang, Mengyun and Huang, Xiangqi and Gan, Xiang and Yang, Yong",
  title="Transferable Adversarial Perturbations",
  booktitle=ECCV,
  year="2018",
  publisher="Springer International Publishing",
  address="Cham",
  pages="471--486",
  isbn="978-3-030-01264-9",
  doi="10.1007/978-3-030-01264-9_28"
}

@INPROCEEDINGS{dong2019evading,
  author={Dong, Yinpeng and Pang, Tianyu and Su, Hang and Zhu, Jun},
  booktitle=CVPR, 
  title={Evading Defenses to Transferable Adversarial Examples by Translation-Invariant Attacks}, 
  year={2019},
  volume={},
  number={},
  pages={4307-4316},
  doi={10.1109/CVPR.2019.00444}
}

@InProceedings{wang2021dual,
  author    = {Wang, Jiakai and Liu, Aishan and Yin, Zixin and Liu, Shunchang and Tang, Shiyu and Liu, Xianglong},
  title     = {Dual Attention Suppression Attack: Generate Adversarial Camouflage in Physical World},
  booktitle = CVPR,
  month     = jun,
  year      = {2021},
  pages     = {8565-8574},
  doi       = {10.1109/cvpr46437.2021.00846}
}

@INPROCEEDINGS{zhang2022investigating,
  author={Zhang, Chaoning and Benz, Philipp and Karjauv, Adil and Cho, Jae Won and Zhang, Kang and Kweon, In So},
  booktitle={2022 IEEE/CVF Conference on Computer Vision and Pattern Recognition (CVPR)}, 
  title={Investigating Top-$k$ White-Box and Transferable Black-box Attack}, 
  year={2022},
  pages={15064-15073},
  doi={10.1109/CVPR52688.2022.01466}
}

@article{he2023improving,
  author = {Xianglong He and Yuezun Li and Haipeng Qu and Junyu Dong},
  title = {Improving transferable adversarial attack via feature-momentum},
  journal = {Computers \& Security},
  volume = {128},
  pages = {103135},
  year = {2023},
  issn = {0167-4048},
  doi = {10.1016/j.cose.2023.103135}
}

@article{gu2024survey,
  author={Jindong Gu and Xiaojun Jia and Pau de Jorge and Wenqian Yu and Xinwei Liu and Avery Ma and Yuan Xun and Anjun Hu and Ashkan Khakzar and Zhijiang Li and Xiaochun Cao and Philip Torr},
  title={A Survey on Transferability of Adversarial Examples Across Deep Neural Networks},
  journal={Transactions on Machine Learning Research},
  issn={2835-8856},
  year={2024},
  url={https://openreview.net/forum?id=AYJ3m7BocI}
}

\appendix

\section{Far-field integrals}\label{appendix:components}

\subsection{Single reflections}

\begin{equation}
\begin{aligned}
    &\begin{cases}
        N_1^\theta = 0, \\ 
        N_1^\phi = \frac{2A^t}{Z_0} \cos\theta' 
            \iint_{\mathcal{A}^i_1} e^{\frac{4j\pi}{\lambda} \left( x \cos\phi' + y \sin\phi' \right) \sin\theta'} \dx\dy.
    \end{cases} \\
    &\begin{cases}
        N_2^\theta = 0, \\
        N_2^\phi = \frac{2A^t}{Z_0} \sin\theta'\cos\phi' 
            \iint_{\mathcal{A}^i_2} e^{\frac{4j\pi}{\lambda} \left( y \sin\theta'\sin\phi' + z \cos\theta' \right)} \dy\dz.
    \end{cases} \\
    &\begin{cases}
        N_3^\theta = 0, \\ 
        N_3^\phi = \frac{2A^t}{Z_0} \sin\theta'\sin\phi' 
        \iint_{\mathcal{A}^i_3} e^{\frac{4j\pi}{\lambda} \left( x \sin\theta'\cos\phi' + z \cos\theta' \right)} \dx\dz.
    \end{cases}
\end{aligned}
\end{equation}

\subsection{Double reflections}
\begin{equation}
\begin{aligned}
    &\begin{cases}
    N_{12}^\theta = -\frac{A^t}{Z_0} \sin(\theta')\sin(\phi') \iint_{\mathcal{A}^i_{12}} e^{\frac{4j\pi}{\lambda} y \sin(\theta')\sin(\phi') } \dy\dz, \\
    N_{12}^\phi = -\frac{2A^t}{Z_0} \sin(\theta')\cos(\phi')\iint_{\mathcal{A}^i_{12}} e^{\frac{4j\pi}{\lambda} y \sin(\theta') \sin(\phi')} \dy\dz.
    \end{cases} \\
    &\begin{cases}
    N_{13}^\theta = \frac{4A^t}{Z_0} \sin(2\theta')\cos(\phi') \iint_{\mathcal{A}^i_{13}} e^{\frac{4j\pi}{\lambda} x \sin(\theta')\cos(\phi') } \dx \dz, \\
    N_{13}^\phi = -\frac{2A^t}{Z_0} \sin(\theta')\sin(\phi')\iint_{\mathcal{A}^i_{13}} e^{\frac{4j\pi}{\lambda} x \sin(\theta') \cos(\phi')} \dx \dz.
    \end{cases} \\
    &\begin{cases}
    N_{21}^\theta = -\frac{4A^t}{Z_0} \cos(\theta')^2\sin(\phi')\cos(\phi') \iint_{\mathcal{A}^i_{21}} e^{\frac{4j\pi}{\lambda} y \sin(\theta')\sin(\phi')} \dx\dy, \\
    N_{21}^\phi = -\frac{2A^t}{Z_0} \cos(\theta')\cos(2\phi')\iint_{\mathcal{A}^i_{21}} e^{\frac{4j\pi}{\lambda} y \sin(\theta') \sin(\phi')} \dx\dy.
    \end{cases} \\
    &\begin{cases}
    N_{23}^\theta = -\frac{4A^t}{Z_0} \sin(\theta')\cos(\theta')\cos(\phi')\iint_{\mathcal{A}^i_{23}} e^{\frac{4j\pi}{\lambda} z \cos(\theta')} \dx\dz, \\
    N_{23}^\phi = \frac{2A^t}{Z_0} \sin(\theta')\sin(\phi')\iint_{\mathcal{A}^i_{23}} e^{\frac{4j\pi}{\lambda} z \cos(\theta')} \dx\dz.
    \end{cases} \\
    &\begin{cases}
    N_{31}^\theta = \frac{4A^t}{Z_0} \cos(\theta')^2\sin(\phi')\cos(\phi') \iint_{\mathcal{A}^i_{31}} e^{\frac{4j\pi}{\lambda} x \sin(\theta')\cos(\phi')} \dx\dy, \\
    N_{31}^\phi = \frac{2A^t}{Z_0} \cos(\theta')\cos(2\phi')\iint_{\mathcal{A}^i_{31}} e^{\frac{4j\pi}{\lambda} x \sin(\theta') \cos(\phi')} \dx\dy.
    \end{cases} \\
    &\begin{cases}
    N_{32}^\theta = \frac{4A^t}{Z_0} \sin(\theta')\cos(\theta')\sin(\phi') \iint_{\mathcal{A}^i_{32}} e^{\frac{4j\pi}{\lambda} z \cos(\theta')\sin(\phi') } \dy\dz, \\
    N_{32}^\phi = \frac{2A^t}{Z_0} \sin(\theta')\cos(\phi')\iint_{\mathcal{A}^i_{32}} e^{\frac{4j\pi}{\lambda} z \cos(\theta')} \dy\dz.
    \end{cases}
\end{aligned}
\end{equation}

\subsection{Triple reflections}

\begin{equation}
\begin{aligned}
    &\begin{cases}
        N_{123}^\theta = 0, \\
        N_{123}^\phi = -\frac{2A^t}{Z_0} \sin(\theta') \sin(\phi') \iint_{\mathcal{A}^i_{123}} \dx\dz.
    \end{cases} \\
    &\begin{cases}
        N_{132}^\theta = 0, \\
        N_{132}^\phi = -\frac{2A^t}{Z_0} \sin(\theta') \cos(\phi') \iint_{\mathcal{A}^i_{132}} \dy\dz.
    \end{cases} \\
    &\begin{cases}
         N_{213}^\theta = 0, \\
        N_{213}^\phi = -\frac{2A^t}{Z_0} \sin(\theta') \sin(\phi') \iint_{\mathcal{A}^i_{213}} \dx\dz.
    \end{cases} \\
    &\begin{cases}
         N_{231}^\theta = 0, \\
        N_{231}^\phi = -\frac{2A^t}{Z_0} \cos(\theta') \iint_{\mathcal{A}^i_{231}} \dx\dy.
    \end{cases} \\
    &\begin{cases}
        N_{312}^\theta = 0, \\
        N_{312}^\phi = -\frac{2A^t}{Z_0} \sin(\theta') \cos(\phi') \iint_{\mathcal{A}^i_{312}} \dy\dz.
    \end{cases} \\
    &\begin{cases}
        N_{321}^\theta = 0, \\
        N_{321}^\phi = -\frac{2A^t}{Z_0}\cos(\theta')\iint_{\mathcal{A}^i_{321}} \dx\dy.
    \end{cases}
\end{aligned}
\end{equation}

\end{document}